\documentclass[%
 reprint,
superscriptaddress,
nofootinbib,
 amsmath,amssymb,
 onecolumn,
prd,
]{revtex4-2}

\usepackage{graphicx}
\usepackage{dcolumn}
\usepackage{bm}
\usepackage[colorlinks]{hyperref}

\usepackage{color}
\usepackage[caption=false]{subfig}

\usepackage{empheq}

\renewcommand{\vec}[1]{\boldsymbol{#1}}

\newcommand{\Mpch}{h^{-1}\mathrm{Mpc}}
\newcommand{\hMpc}{h\,\mathrm{Mpc}^{-1}}

\newcommand{\av}[1]{\left\langle{#1}\right\rangle}

\newcommand{\vk}{\vec k}

\newcommand{\vs}{\vec s}

\renewcommand{\vr}{\vec r}

\renewcommand{\L}{\Lambda}
\renewcommand{\P}{\mathcal{P}}
\newcommand{\hR}{\hat{\vec R}}
\newcommand{\hr}{\hat{\vec r}}
\newcommand{\hn}{\hat{\vec n}}
\newcommand{\hk}{\hat{\vec k}}
\newcommand{\tjo}[3]{\begin{pmatrix} {#1} & {#2} & {#3}\\ 0 & 0 & 0\end{pmatrix}}
\newcommand{\tj}[6]{\begin{pmatrix} {#1} & {#2} & {#3}\\ {#4} & {#5} & {#6}\end{pmatrix}}

\usepackage{color}
\def\beq{\begin{eqnarray}}
\def\eeq{\end{eqnarray}}

\definecolor{darkgreen}{RGB}{0,120,0}

\begin{document}


\title{A First Detection of the Connected 4-Point Correlation Function of Galaxies Using the BOSS CMASS Sample}

\author{Oliver H.\,E. Philcox}
\email{ohep2@cantab.ac.uk}
\affiliation{Department of Astrophysical Sciences, Princeton University,\\ Princeton, NJ 08540, USA}%
\affiliation{School of Natural Sciences, Institute for Advanced Study, 1 Einstein Drive,\\ Princeton, NJ 08540, USA}
\author{Jiamin Hou}%
\affiliation{Department of Astronomy, University of Florida,\\
211 Bryant Space Science Center, Gainesville, FL 32611, USA}%
\author{Zachary Slepian}%
\affiliation{Department of Astronomy, University of Florida,\\
211 Bryant Space Science Center, Gainesville, FL 32611, USA}%
\affiliation{Lawrence Berkeley National Laboratory,\\ 1 Cyclotron Road, Berkeley, CA 94720, USA}



\begin{abstract}
We present an $8.1\sigma$ detection of the non-Gaussian 4-Point Correlation Function (4PCF) using a sample of $N_{\rm g} \approx 8\times 10^5$ galaxies from the BOSS CMASS dataset. Our measurement uses the $\mathcal{O}(N_{\rm g}^2)$ NPCF estimator of Philcox \textit{et al.} (2021), including a new modification to subtract the disconnected 4PCF contribution (arising from the product of two 2PCFs) at the estimator level. This approach is unlike previous work and ensures that our signal is a robust detection of gravitationally-induced non-Gaussianity. The estimator is validated with a suite of lognormal simulations, and the analytic form of the disconnected contribution is discussed. Due to the high dimensionality of the 4PCF, data compression is required; we use a signal-to-noise-based scheme calibrated from theoretical covariance matrices to restrict to $\sim$\,$100$ basis vectors. The compression has minimal impact on the detection significance and facilitates traditional $\chi^2$-like analyses using a suite of mock catalogs. The significance is stable with respect to different treatments of noise in the sample covariance (arising from the limited number of mocks), but decreases to $4.7\sigma$ when a minimum galaxy separation of $14\,\Mpch$ is enforced on the 4PCF tetrahedra (such that the statistic can be modelled more easily). The detectability of the 4PCF in the quasi-linear regime implies that it will become a useful tool in constraining cosmological and galaxy formation parameters from upcoming spectroscopic surveys.
\end{abstract}

\maketitle


\section{Introduction}\label{sec: intro}

The detection of a non-Gaussian signature in the early Universe would be a smoking gun for many inflation models \citep[e.g.,][]{2004PhR...402..103B}. Despite a number of searches, no evidence has been found for primordial non-Gaussianity in the Cosmic Microwave Background (CMB) \citep{2020A&A...641A...9P}. Given that its signal in the CMB is found primarily on large, cosmic variance dominated scales, the situation is unlikely to change dramatically in the near future \citep{2019JCAP...02..056A}. With the upcoming influx of data from spectroscopic surveys such as DESI \citep{2016arXiv161100036D} and Euclid \citep{2011arXiv1110.3193L}, it is natural to expect that large-scale structure (LSS) will soon become a key dataset with which to probe such signatures. 

At late times, there is an additional source of non-Gaussianity: structure growth induced by gravitational evolution \citep[e.g.,][]{2002PhR...367....1B}. Far from simply being a contaminant in the search for signatures from early-Universe physics, gravitational non-Gaussianities provide significant information on cosmological parameters such as the growth rate, neutrino masses, and modified gravity parameters \citep[e.g.,][]{2017MNRAS.467..928G,2021JCAP...03..021A,2019JCAP...11..034C,2021JCAP...04..029H}. As such, measuring non-Gaussian statistics such as the higher-order $N$-Point Correlation Functions (NPCFs) or their Fourier-space equivalents (polyspectra) provides a powerful approach with which to enhance cosmological analyses, capturing information shifted out of the Gaussian two-point function due to gravitational evolution \citep{2015PhRvD..92l3522S}. Furthermore, it is vital to understand the late-time contributions to these statistics if we wish to use them to extract constraints on primordial non-Gaussianities \citep{2018MNRAS.478.1341K,2021JCAP...05..015M}.

A number of recent analyses have included the simplest non-Gaussian statistic; the 3-Point Correlation Function (3PCF) \citep{2017MNRAS.468.1070S,2018MNRAS.474.2109S}, or bispectrum \citep{2015MNRAS.451..539G,2018MNRAS.478.4500P,2019MNRAS.484.3713G,2020JCAP...05..005D,2017MNRAS.465.1757G}. This has been facilitated by fast algorithms for computing such statistics \citep[e.g.,][]{2015MNRAS.454.4142S, SE_3PCF_FT,2018MNRAS.478.1468S,2004ApJ...605L..89S,2005NewA...10..569Z,2001misk.conf...71M,2015PhRvD..92h3532S,garcia_2020, sugiyama_2018}, as well as the development of theory models for the galaxy 3PCF \citep{Slepian_2017} and bispectrum \citep[e.g.,][]{2000ApJ...544..597S,2003MNRAS.340..580T,2005MNRAS.361..824G,2008ApJ...672..849M,2017MNRAS.469.2059S,2015JCAP...05..007B,2015JCAP...10..039A}, though the latter is still in its infancy. The additional information contained within the 3PCF has sharpened cosmological parameter constraints \citep{2017MNRAS.465.1757G,2019MNRAS.484L..29G,2019MNRAS.484.3713G}, particularly through the Baryon Acoustic Oscillation (BAO) feature \citep{2017MNRAS.469.1738S,2018MNRAS.478.4500P}, and probed the relative velocity of baryons and dark matter \citep{SE_RV_theory, 2018MNRAS.474.2109S}.

\begin{figure}
    \centering
    \includegraphics[width=0.9\textwidth]{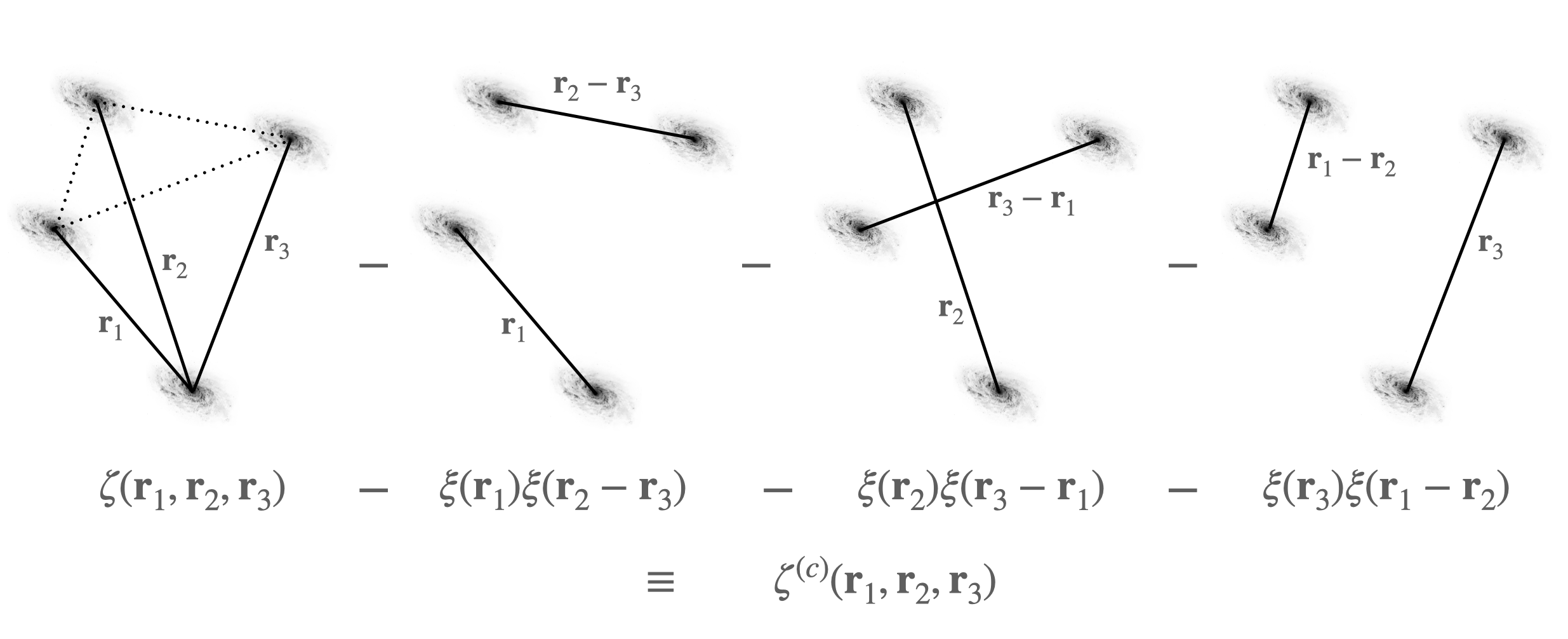}
    \caption{Schematic of the connected galaxy 4PCF. The first cartoon shows the contribution of a single tetrahedron of galaxies to the full 4PCF, $\zeta(\vr_1,\vr_2,\vr_3)$, which is parametrized by the separations of three secondary galaxies from a given primary (at the base of the tetrahedron). The complete statistic can be formed by summing over all such tetrahedra of galaxies; in practice, we use a spherical harmonic decomposition which is considerably more efficient \citep{npcf_algo}. In the second through fourth diagrams, we show the contributions to the \textit{disconnected} 4PCF, arising from products of two 2PCFs, $\xi(\vr)$. The complete statistic may be computed by iterating over pairs of galaxies. The difference between the full and disconnected configurations gives the connected 4PCF, $\zeta^{(\rm c)}(\vr_1,\vr_2,\vr_3)$, which is the quantity of interest in this work.} 
    \label{fig: connected-cartoon}
\end{figure}

In this work, we focus on the next non-Gaussian statistic: the 4-Point Correlation Function (4PCF). This has been scarcely considered in the literature; a handful of works explore its estimation \citep{2005NewA...10..569Z,2019ApJS..242...29S,2019AJ....158..116T}, and, for the matter field, its modelling \citep{2016JCAP...06..052B,2021JCAP...01..015G}, and projected parameter constraints \citep{2021arXiv210403976G}. Besides sharpening constraints on $\Lambda$CDM parameters, the 4PCF can also be used to test new physics such as parity violation \citep{2016PhRvD..94h3503S}; this will be considered in depth in \citep{4pcf_odd}. Unlike the 2PCF and 3PCF, the 4PCF contains both an intrinsic (`connected') four-point function, and a disconnected piece, which depends on the product of two 2PCFs (as sketched in Fig.\,\ref{fig: connected-cartoon}). Given that the latter contains only information degenerate with that in the 2PCF, it is important to separate the two contributions. Here, our goal is to extract only the \textit{connected} 4PCF, unlike previous work \citep{2019ApJS..242...29S}.

Performing a 4PCF analysis comes with a number of challenges. Firstly, estimation of the statistic is non-trivial, with na\"ive approaches having $\mathcal{O}(N_{\rm g}^4)$ complexity when applied to a dataset containing $N_{\rm g}$ galaxies. Here, we apply the recently-proposed $\mathcal{O}(N_{\rm g}^2)$ NPCF estimators \citep{npcf_algo,npcf_generalized} which achieve a significant speed-boost by projecting the statistic onto a separable angular basis obtained from the theory of angular momentum addition \citep{2020arXiv201014418C}. Coupled with a new procedure for removing the disconnected component at the estimator level, this allows the connected isotropic 4PCF to be computed in only a few tens of CPU-hours, including survey geometry corrections. Furthermore, the dimensionality of the 4PCF is large, making traditional mock-based analyses difficult. To ameliorate this, we use the signal-to-noise based compression scheme of \citep{2000ApJ...544..597S}, which greatly reduces the number of bins with minimal impact on the detection significance. A key part of the compression scheme is a smooth approximation to the 4PCF covariance; for this purpose, we make use of analytic NPCF covariances derived in \citep{npcf_cov} under certain simplifying assumptions. This facilitates a traditional $\chi^2$ analysis calibrated using realistic mock catalogs, allowing our 4PCF detection significance to be quantified, albeit under the assumption of a Gaussian likelihood.

\vskip 4pt

The remainder of this paper is structured as follows. \S\ref{sec: background} provides a summary of the 4PCF estimator introduced in \citep{npcf_algo}, before we discuss the modifications required to remove the disconnected 4PCF in \S\ref{sec: connected-estimator}. Using a suite of lognormal simulations and an analytic form for the disconnected 4PCF, the estimator is tested in \S\ref{sec: lognormals}. In \S\ref{sec: techniques}, we outline our data compression scheme, with the data, simulations and analytic covariance being introduced in \S\ref{sec: data}. The results of our analysis are presented in \S\ref{sec: results}, before we conclude in \S\ref{sec: summary}. Appendices \ref{appen: gaussian-deriv}, \ref{appen: rescaled-theory-cov} \& \ref{appen: fully-coupled-covariance} present various useful results and consistency checks. Our main results are displayed in Figs.\,\ref{fig: boss-4pcfs}\,\&\,\ref{fig: boss-4pcf-detection}, respectively showing a selection of the BOSS 4PCF measurements, and the detection significance.

\vskip 4pt

Throughout this work we use the fiducial cosmology $\{\Omega_m = 0.31, \Omega_bh^2 = 0.022, h = 0.676, \sigma_8 = 0.8, n_s = 0.96, \sum m_\nu = 0.06\,\mathrm{eV}\}$ as in \citep{2017MNRAS.466.2242B}; the neutrino sector is modeled as a single massive neutrino. Furthermore, we adopt the Fourier conventions $P(\vk) = \int d\vr\,e^{-i\vk\cdot\vr}\xi(\vr) \leftrightarrow \xi(\vr) = \int_{\vk} P(\vk)e^{i\vk\cdot\vr}\xi(\vr)$ for power spectrum $P(\vk)$ and 2PCF $\xi(\vr)$, denoting $(2\pi)^{-3}\int d\vk \equiv \int_{\vk}$. The main analysis pipeline of this work is available on GitHub.\footnote{\href{https://github.com/oliverphilcox/Parity-Even-4PCF}{github.com/oliverphilcox/Parity-Even-4PCF}}

\section{The Full 4PCF Estimator}\label{sec: background}
We begin with a summary of the isotropic 4PCF estimator of \citep{npcf_algo, npcf_generalized}, including discussion of the relevant angular basis \citep{2020arXiv201014418C}. Note that this estimates both the disconnected and connected 4PCF; the removal of the former piece is described in \S\ref{sec: connected-estimator}. All estimators are implemented in the public \textsc{encore} code, written in \textsc{C++} and \textsc{cuda}.\footnote{\href{https://github.com/oliverphilcox/encore}{github.com/oliverphilcox/encore}.}

\subsection{Idealized Estimator}\label{subsec: full-4pcf-estimator}
The 4PCF is defined as a statistical expectation of four copies of the density field, $\delta$:
\beq\label{eq: 4pcf-def}
    \zeta(\vr_1,\vr_2,\vr_3) \equiv \av{\delta(\vs)\delta(\vs+\vr_1)\delta(\vs+\vr_2)\delta(\vs+\vr_3)}.
\eeq
By definition, the \textit{isotropic} 4PCF depends only on the shape of the quadrilateral defined by the separation vectors $\{\vr_1,\vr_2,\vr_3\}$ (as in Fig.\,\ref{fig: connected-cartoon}), and not on its spatial position or orientation. As shown in \citep{2020arXiv201014418C}, any isotropic function of $(N-1)$ coordinates in 3D can be expanded in terms of the following basis vectors:
\beq\label{eq: general-basis}
    \P_{\L}(\hr_1,\ldots,\hr_{N-1}) &=& \sum_{M}C_{\L}^{M}\;Y_{\ell_1}^{m_1}(\hr_1)\cdots Y_{\ell_{N-1}}^{m_{N-1}}(\hr_{N-1}),
\eeq
where $\L$ and $M$ are sets of total and projected angular momentum indices respectively, $C_{\L}^M$ is a coupling matrix, and $Y_{\ell}^m(\hr)$ is a spherical harmonic.\footnote{These are invariant under joint rotations of $\hR\equiv\{\hr_1,\ldots,\hr_{N-1}\}$, \textit{i.e.} $\P_{\L}(\mathcal{R}\hR) = \P_\L(\hR)$ for arbitrary rotation operator $\mathcal{R}$.} Since the dependence of \eqref{eq: general-basis} on the unit vectors $\hr_i$ is separable, the $\mathcal{P}_{\Lambda}$ functions offer a convenient basis for efficient computation of the isotropic NPCFs \citep{npcf_algo}. For $N=4$, the basis becomes
\beq\label{eq: N=4-basis}
    \boxed{\P_{\ell_1\ell_2\ell_3}(\hr_1,\hr_2,\hr_3) = (-1)^{\ell_1+\ell_2+\ell_3}\sum_{m_1=-\ell_1}^{\ell_1}\sum_{m_2=-\ell_2}^{\ell_2}\sum_{m_3=-\ell_3}^{\ell_3}\tj{\ell_1}{\ell_2}{\ell_3}{m_1}{m_2}{m_3}Y_{\ell_1}^{m_1}(\hr_1)Y_{\ell_2}^{m_2}(\hr_2)Y_{\ell_3}^{m_3}(\hr_3),}
\eeq
where the $2\times3$ matrix is a Wigner 3-$j$ symbol \citep[\S34.3]{nist_dlmf} and the basis depends on three angular momentum indices $\{\ell_1,\ell_2,\ell_3\}$, which (a) can take only non-negative integer values, and (b) must satisfy the triangle condition $|\ell_1-\ell_2|\leq \ell_3\leq \ell_1+\ell_2$. \eqref{eq: N=4-basis} is closely related to the tripolar spherical harmonic (TriPoSH) basis functions of zero total angular momentum \citep{1988qtam.book.....V}. 
The isotropic 4PCF can be expanded in this basis as 
\beq
    \zeta(\vr_1,\vr_2,\vr_3) = \sum_{\ell_1\ell_2\ell_3}\zeta_{\ell_1\ell_2\ell_3}(r_1,r_2,r_3)\P_{\ell_1\ell_2\ell_3}(\hr_1,\hr_2,\hr_3);
\eeq
by orthogonality, the coefficients (hereafter denoted `multiplets') can be obtained via
\beq\label{eq: 4pcf-multiplets-extraction}
    \zeta_{\ell_1\ell_2\ell_3}(r_1,r_2,r_3) &=& \int d\hr_1 d\hr_2 d\hr_3\,\zeta(\vr_1,\vr_2,\vr_3)\,\P^*_{\ell_1\ell_2\ell_3}(\hr_1,\hr_2,\hr_3).
\eeq
In this work, we assume the 4PCF to be invariant under the parity transform $\{\vr_1,\vr_2,\vr_3\}\rightarrow \{-\vr_1,-\vr_2,-\vr_3\}$ (\textit{i.e.} parity-even); this implies that the multiplets $\zeta_{\ell_1\ell_2\ell_3}$ are non-zero only for even $\ell_1+\ell_2+\ell_3$ \citep{npcf_algo}. Furthermore, the condition ensures that both the basis functions and the 4PCF coefficients are real. The parity-odd case will be discussed in \citep{4pcf_odd}; standard model physics is not expected to generate such a contribution on large scales.

Via the ergodic principle, the full 4PCF may be estimated as a spatial integral over four copies of $\delta$, assuming statistical homogeneity:
\beq\label{eq: 4pcf-basic-estimator}
    \hat\zeta(\vr_1,\vr_2,\vr_3) = \frac{1}{V}\int d\vs\,\delta(\vs)\delta(\vs+\vr_1)\delta(\vs+\vr_2)\delta(\vs+\vr_3),
\eeq
where $V$ is the survey volume. In combination with \eqref{eq: 4pcf-multiplets-extraction}, this gives an estimator for the 4PCF multiplets in radial bins $\{a,b,c\}$:
\beq\label{eq: 4pcf-multiplet-estimator}
    \hat\zeta^{abc}_{\ell_1\ell_2\ell_3} = \frac{1}{V}\int d\vs\,d\vr_1 d\vr_2 d\vr_3\,\delta(\vs)\delta(\vs+\vr_1)\delta(\vs+\vr_2)\delta(\vs+\vr_3)\,\P^*_{\ell_1\ell_2\ell_3}(\hr_1,\hr_2,\hr_3)\frac{\Theta^a(r_1)}{v_a}\frac{\Theta^b(r_2)}{v_b}\frac{\Theta^c(r_3)}{v_c},
\eeq
where the binning functions $\Theta^u(r)$ are unity if $r=|\vr|$ is in bin $u$ and zero else, and $v_u$ is the bin volume. The estimator can be factorized:
\beq\label{eq: 4pcf-multiplet-factorized-estimator}
    \hat\zeta^{abc}_{\ell_1\ell_2\ell_3} &=& \frac{1}{V}\int d\vs\,
    \delta(\vs)\sum_{m_1m_2m_3}\tj{\ell_1}{\ell_2}{\ell_3}{m_1}{m_2}{m_3}\left[\frac{1}{v_a}\int d\vr_1\,\delta(\vs+\vr_1)Y_{\ell_1}^{m_1}(\hr_1)\Theta^a(r_1)\right]\\\nonumber
    &&\,\times\,\left[\frac{1}{v_b}\int d\vr_2\,\delta(\vs+\vr_2)Y_{\ell_2}^{m_2}(\hr_2)\Theta^b(r_2)\right]\left[\frac{1}{v_c}\int d\vr_3\,\delta(\vs+\vr_3)Y_{\ell_3}^{m_3}(\hr_3)\Theta^c(r_3)\right]\nonumber
\eeq
\beq
    \Rightarrow \boxed{  \hat\zeta^{abc}_{\ell_1\ell_2\ell_3} = \frac{1}{V}\int d\vs\,
    \delta(\vs)\sum_{m_1m_2m_3}\tj{\ell_1}{\ell_2}{\ell_3}{m_1}{m_2}{m_3}a^a_{\ell_1m_1}(\vs)\,a^b_{\ell_2m_2}(\vs)\,a^c_{\ell_3m_3}(\vs),}
\eeq
inserting \eqref{eq: N=4-basis} and defining the coefficients $a_{\ell m}^u(\vs)$ in the final line (using the conjugate properties of the basis \citep{2020arXiv201014418C}). 

For discrete data, such as that from galaxy surveys, the 4PCF multiplets of $N_{\rm g}$ galaxies may be computed in $\mathcal{O}(N_{\rm g}^2)$ time by first computing the $a_{\ell m}^u$ functions at the location of each primary galaxy $\vs$, then combining these via \eqref{eq: 4pcf-multiplet-factorized-estimator}. Given some field $\mathcal{Z}$ estimated from $N_{\rm p}$ particles located at $\{\vs_i\}$ with weights $\{w_i\}$, thie estimator takes the explicit form:
\beq\label{eq: discrete-4pcf-counts}
    \mathcal{Z}^{abc}_{\ell_1\ell_2\ell_3} &=& \frac{1}{V}\sum_{i=1}^{N_{\rm p}}w_i\sum_{m_1m_2m_3}\tj{\ell_1}{\ell_2}{\ell_3}{m_1}{m_2}{m_3}a^a_{\ell_1m_1}(\vs_i)\,a^b_{\ell_2m_2}(\vs_i)\,a^c_{\ell_3m_3}(\vs_i),\\\nonumber
    a_{\ell m }^u(\vs) &=& \sum_{j=1}^{N_{\rm p}}w_j\, Y_{\ell}^{m}(\widehat{\vr_j-\vs})\,\Theta^u(|\vr_j-\vs|),
\eeq
where the $a_{\ell m}^u(\vs)$ summation is only over particles whose separations from $\vs$ lie within  bin $u$. Strictly, this factorization is correct only if the bins $a$, $b$ and $c$ are not coincident (due to shot noise effects, cf.\,\citep{npcf_algo}); in practice we ensure this by enforcing $a<b<c$.

\subsection{Edge Correction}\label{subsec: full-edge-correction}
In practice, we do not have access to the overdensity field $\delta$ directly, and must work instead with a set of data and random particles. Given weights $\{w_i^D\}$ and $\{w_j^R\}$, we define
\beq\label{eq: data-minus-randoms}
    N(\vr) \equiv D(\vr) - \alpha\,R(\vr), \quad D(\vr) \equiv \sum_{i=1}^{N_D} w^D_i\,\delta_\mathrm{D}(\vr-\vr_i), \quad R(\vr) \equiv \sum_{j=1}^{N_R} w^R_j\,\delta_\mathrm{D}(\vr-\vr_j), \quad \alpha \equiv \sum_{i=1}^{N_{\rm g}}w^D_i\big{/}\sum_{j=1}^{N_R}w^R_j,
\eeq
allowing estimation of the full 4PCF via a generalization of \citep{1993ApJ...412...64L}:
\beq
    \hat\zeta(\vr_1,\vr_2,\vr_3) &=& \frac{\mathcal{N}(\vr_1,\vr_2,\vr_3)}{\mathcal{R}(\vr_1,\vr_2,\vr_3)}.
\eeq
Here, $\mathcal{N}$ and $\mathcal{R}$ represent 4PCF counts \eqref{eq: discrete-4pcf-counts} involving four $N$ and four $R$ fields respectively.\footnote{For the 2PCF (3PCF), these are generally referred to as $NN$ and $RR$ ($NNN$ and $RRR$) and often expanded as $NN=DD-2DR+RR$ ($NNN = DDD-3DDR-3DRR+RRR$). We do not perform such an expansion here, opting instead to compute multiplets of the combined `data-minus-randoms' fields for efficiency. This is easily done by multiplying the random weights by a factor $-\alpha$, and concatenating the data and randoms.} Using \eqref{eq: discrete-4pcf-counts}, we may compute the multiplets $\mathcal{N}_{\ell_1\ell_2\ell_3}$ and $\mathcal{R}_{\ell_1\ell_2\ell_3}$, which can be combined to form the binned 4PCF estimates via the \textit{edge correction} equation:
\beq\label{eq: edge-correction-eq}
    \boxed{\zeta_{\ell_1\ell_2\ell_3}^{abc} = \sum_{\ell_1'\ell_2'\ell_3'}\left[\mathsf{M}^{-1}\right]_{\ell_1\ell_2\ell_3}^{\ell_1'\ell_2'\ell_3',abc}\;\frac{\mathcal{N}_{\ell_1'\ell_2'\ell_3'}^{abc}}{\mathcal{R}_{000}^{abc}}}
\eeq
\citep[cf.\,][]{npcf_algo}. This involves a coupling matrix, $\mathsf{M}$, with elements
\beq
    \mathsf{M}^{\ell_1'\ell_2'\ell_3',abc}_{\ell_1\ell_2\ell_3} &=& (4\pi)^{-3/2}\sum_{L_1L_2L_3}\frac{\mathcal{R}_{L_1L_2L_3}^{abc}}{\mathcal{R}_{000}^{abc}} \left[\prod_{i=1}^3\sqrt{(2\ell_i+1)(2L_i+1)(2\ell_i'+1)}\right]\begin{Bmatrix}\ell_1 &L_1 &\ell_1' \\ \ell_2 & L_2 & \ell_2' \\ \ell_3 & L_3 & \ell_3'\end{Bmatrix}\nonumber\\
    &&\,\times \tjo{\ell_1}{L_1}{\ell_1'}\tjo{\ell_2}{L_2}{\ell_2'}\tjo{\ell_3}{L_3}{\ell_3'},
\eeq
where the $3 \times 3$ matrix in curly brackets is a Wigner 9-$j$ symbol \citep[e.g.,][]{1988qtam.book.....V}. In practice, the matrix elements are straightforward to compute from the random counts $\mathcal{R}_{\ell_1\ell_2\ell_3}$. Given the set of all $\{\ell_1,\ell_2,\ell_3\}$ triplets, $\mathsf{M}$ may be written as an invertible two-dimensional matrix. In the limit of a uniform periodic box geometry with unweighted particles, the random counts are known precisely:
\beq\label{eq: periodic-4pcf}
    \left.\mathcal{R}_{\ell_1\ell_2\ell_3}^{abc}\right|_\mathrm{periodic} = \frac{\bar{n}^4}{(4\pi)^{3/2}}\, \delta^\mathrm{K}_{\ell_10}\delta^\mathrm{K}_{\ell_20}\delta^\mathrm{K}_{\ell_30} \quad \Rightarrow \quad  \left.\zeta_{\ell_1\ell_2\ell_3}^{abc}\right|_\mathrm{Periodic} = \frac{(4\pi)^{3}}{\bar{n}^{4}}\,\mathcal{N}_{\ell_1\ell_2\ell_3}^{abc}
\eeq
where $\delta^\mathrm{K}$ is the Kronecker delta, and $\bar{n}$ is the number density. In practice, the non-uniform survey geometry generates higher-order $\mathcal{R}$ multiplets, thus the full form of \eqref{eq: edge-correction-eq} is required. The coupling matrix $\mathsf{M}$ is usually found to be close to tridiagonal \citep{npcf_algo}, such that, when measuring data and random multiplets up to $\ell_i=\ell_\mathrm{max}$, the output NPCFs are accurate up to $\ell_i=(\ell_\mathrm{max}-1)$ (since the contributions from higher multiplets are small).

\section{The Connected 4PCF Estimator}\label{sec: connected-estimator}
Here, we discuss an improved estimator that removes the disconnected 4PCF arising from the product of two 2PCFs. An alternative approach would be to retain the disconnected component and account for it via a theory model (described in Appendix \ref{appen: gaussian-deriv}); this approach is disfavored since it requires an accurate model of the two-point function across a large range of scales.

\subsection{Idealized Disconnected Estimator}\label{subsec: disconnected-expectation}
Averaging over realizations of the density field $\delta$, the product of four density fields gives
\beq
    \av{\delta(\vs)\delta(\vs+\vr_1)\delta(\vs+\vr_2)\delta(\vs+\vr_3)} &=& \av{\delta(\vs)\delta(\vs+\vr_1)\delta(\vs+\vr_2)\delta(\vs+\vr_3)}_c \\\nonumber
    &&\,\quad\,+\,\left[\av{\delta(\vs)\delta(\vs+\vr_1)}\av{\delta(\vs+\vr_2)\delta(\vs+\vr_3)} + \text{2 perms.}\right] \\\nonumber
    &\equiv& \zeta^{(\rm c)}(\vr_1,\vr_2,\vr_3) + \left[\xi(\vr_1)\xi(\vr_2-\vr_3) + \text{2 perms.}\right],
\eeq
assuming statistical homogeneity (but not isotropy) and denoting connected pieces by `$c$'. This decomposition is visualized in Fig.\,\ref{fig: connected-cartoon}. Given the standard 2PCF estimator,
\beq\label{eq: 2pcf-estimator}
    \hat\xi(\vr) = \frac{1}{V}\int d\vs\,\delta(\vs)\delta(\vs+\vr),
\eeq
we can write a simple estimator for the disconnected 4PCF:\footnote{Alternatively, this can be obtained from \eqref{eq: 4pcf-multiplet-estimator} by replacing the product of four density fields with two 2PCF estimators and separating the resulting expression.}
\beq
    \hat{\zeta}^{\mathrm{(disc)}}(\vr_1,\vr_2,\vr_3) &=& \hat{\xi}(\vr_1)\hat{\xi}(\vr_2-\vr_3) + \text{2 perms.}\\\nonumber
    &=& \left[\frac{1}{V}\int d\vs\,\delta(\vs)\delta(\vs+\vr_1)\right]\left[\frac{1}{V}\int d\vs'\,\delta(\vs'+\vr_2)\delta(\vs'+\vr_3)\right] + \text{2 perms.},
\eeq
before binning. This is unbiased, such that $\av{\hat\zeta^{(\mathrm{disc})}(\vr_1,\vr_2,\vr_3)} = \zeta^{(\mathrm{disc})}(\vr_1,\vr_2,\vr_3)$.\footnote{Strictly this is true only in the limit of infinite volume $V$. Since corrections scale as $r_c^3/V$ for typical correlation length $r_c$\,$\sim$\,$100\,\Mpch$, this is of little importance for current and future surveys.}

Projecting onto the isotropic basis functions \eqref{eq: N=4-basis}, we obtain an expression analogous to \eqref{eq: 4pcf-multiplet-estimator}:
\beq\label{eq: disconnected-estimator-tmp}
    \hat\zeta^{(\mathrm{disc})}_{\ell_1\ell_2\ell_3}(r_1,r_2,r_3) &\equiv& \left[\frac{1}{V}\int d\vs\, d\hr_1\,\delta(\vs)\delta(\vs+\vr_1)\right]\left[\frac{1}{V}\int d\vs'd\hr_2d\hr_3\,\delta(\vs'+\vr_2)\delta(\vs'+\vr_3)\right]\P^*_{\ell_1\ell_2\ell_3}(\hr_1,\hr_2,\hr_3) + \text{2 perms.}\nonumber\\ &=&(-1)^{\ell_1+\ell_2+\ell_3}\sum_{m_1m_2m_3}\tj{\ell_1}{\ell_2}{\ell_3}{m_1}{m_2}{m_3}\left[\frac{1}{V}\int d\vs\,d\hr_1\,\delta(\vs)\delta(\vs+\vr_1)Y^*_{\ell_1m_1}(\hr_1)\right]\nonumber\\
    &&\,\times\,\left[\frac{1}{V}\int d\vs'd\hr_2d\hr_3\,\delta(\vs'+\vr_2)\delta(\vs'+\vr_3)Y^*_{\ell_2m_2}(\hr_2)Y^*_{\ell_3m_3}(\hr_{3})\right] + \text{2 perms.}
\eeq
Defining
\beq\label{eq: xi-lm-lmlmm-def}
    \hat\xi_{\ell m}(r) &\equiv& \frac{1}{V}\int d\vs\,d\hr\,\delta(\vs)\delta(\vs+\vr)Y^*_{\ell m}(\hr), \\\nonumber
    \hat\xi_{\ell m \ell'm'}(r,r') &\equiv& \frac{1}{V}\int d\vs' d\hr\, d\hr'\,\delta(\vs'+\vr)\delta(\vs'+\vr')Y^*_{\ell m}(\hr)Y^*_{\ell'm'}(\hr'),
\eeq
\eqref{eq: disconnected-estimator-tmp} can be written
\beq\label{eq: disconnected-4pcf-estimator}
    \hat\zeta_{{\ell_1\ell_2\ell_3}}^{(\mathrm{disc})}(r_1,r_2,r_3) &=& (-1)^{\ell_1+\ell_2+\ell_3}\sum_{m_1m_2m_3}\tj{\ell_1}{\ell_2}{\ell_3}{m_1}{m_2}{m_3}\hat\xi_{\ell_1m_1}(r_1)\hat\xi_{\ell_2 m_2\ell_3m_3}(r_2,r_3)+\text{2 perms.}.
\eeq
Combining with the full 4PCF estimator \eqref{eq: 4pcf-multiplet-factorized-estimator}, we obtain an unbiased estimator for the \textit{connected} 4PCF as
\beq\label{eq: connected-4pcf-estimator}
    \boxed{\hat\zeta_{\ell_1\ell_2\ell_3}^{(\rm c)}(r_1,r_2,r_3) = \hat\zeta_{\ell_1\ell_2\ell_3}(r_1,r_2,r_3)-\hat\zeta^{(\mathrm{disc})}_{\ell_1\ell_2\ell_3}(r_1,r_2,r_3).}
\eeq
Incorporating radial binning in this expression is straightforward (cf.\,\ref{eq: 4pcf-multiplet-estimator}).

\subsection{Computation of $\xi_{\ell m}(r)$ and $\xi_{\ell m\ell'm'}(r,r')$}\label{subsec: discon-edge-correction}
To estimate the connected 4PCF of discrete data, we must compute the multipoles $\xi_{\ell m}(r)$ and $\xi_{\ell m\ell'm'}(r,r')$ from the set of data and random particles. Computation of these is analogous to that of the isotropic 2PCF and 3PCF, and has $\mathcal{O}(N_{\rm g}^2)$ complexity.

For $\hat{\xi}_{\ell m}(r)$, we may write
\beq\label{eq: sph-pair-counts}
    \hat\xi_{\ell m}(r) = \frac{1}{Vv(r)}\int d\vs\,d\hr\,\delta(\vs)\delta(\vs+\vr)Y^*_{\ell m}(\hr) \equiv \frac{1}{V}\int d\vs\,\delta(\vs)\,a^*_{\ell m}(\vs;r),
\eeq
using the $a_{\ell m}(\vs; r)$ quantities defined for estimation of the full NPCF in \eqref{eq: 4pcf-multiplet-factorized-estimator}, and including a (continuous) bin volume $v(r)$. To compute this from discrete data, we use the standard Landy-Szalay estimator for $\hat{\xi}(\vr)$ \citep{1993ApJ...412...64L}, giving:
\beq
    \hat\xi(\vr) = \frac{NN(\vr)}{RR(\vr)} &\Rightarrow& \sum_{\ell m}\hat\xi_{\ell m}(r)Y_{\ell m}(\hr) = \frac{\sum_{LM}NN_{LM}(r)Y_{LM}(\hr)}{\sum_{\ell'm'}RR_{\ell'm'}(r)Y_{\ell'm'}(\hr)},
\eeq
expanding $\hat\xi(\vr)$, $NN(\vr)$, and $RR(\vr)$ in the (complete) basis of spherical harmonics in the second equation. Following some algebra, we obtain
the edge correction equation for $\xi_{\ell m}(r)$, analogous to \eqref{eq: edge-correction-eq}:
\beq\label{eq: xi-lm-edge-correction}
    \boxed{\hat\xi_{\ell m}(r) = \sum_{LM}\left[\sum_{\ell'm'}\frac{RR_{\ell'm'}(r)}{RR_{00}(r)}\mathcal{G}_{\ell\ell'L}^{mm'-M}(-1)^M\right]^{-1}\frac{NN_{LM}(r)}{RR_{00}(r)}, }
\eeq
where $\mathcal{G}^{mm'(-M)}_{\ell\ell'L}$ is the Gaunt integral, defined by
\beq
    \mathcal{G}_{\ell_1\ell_2\ell_3}^{m_1m_2m_3} \equiv \sqrt{\frac{(2\ell_1+1)(2\ell_2+1)(2\ell_3+1)}{4\pi}}\tj{\ell_1}{\ell_2}{\ell_3}{m_1}{m_2}{m_3}\tjo{\ell_1}{\ell_2}{\ell_3}
\eeq
\citep[\S34.3]{nist_dlmf}. \eqref{eq: xi-lm-edge-correction} is straightforward to compute given the $NN_{LM}$ and $RR_{LM}$ pair counts:
\beq\label{eq: NN,RR-def}
    NN_{\ell m}(r) = \int \frac{d\vs}{V}\,N(\vs)\,a^*_{\ell m}[N](\vs;r), \quad RR_{\ell m}(r) = \int \frac{d\vs}{V}\,R(\vs)\,a^*_{\ell m}[R](\vs;r),
\eeq
where $a_{\ell m}[X](\vs;r)$ are the usual $a_{\ell m}$ coefficients with $\delta$ replaced by $X$. For a periodic survey geometry with $R(\vr)\equiv\bar n$, these expressions simplify 
\beq\label{eq: periodic-xi-lm}
    \left.RR_{\ell m}(r)\right|_\mathrm{periodic} = \frac{\bar n^2}{\sqrt{4\pi}}\,\delta^\mathrm{K}_{\ell 0}\delta^\mathrm{K}_{m0} \qquad \Rightarrow \qquad  \left.\hat\xi_{\ell m}(r)\right|_\mathrm{periodic} &=& \frac{4\pi}{\bar n^2}NN_{\ell m}(r).
\eeq

$\hat\xi_{\ell m\ell'm'}(r,r')$ may be computed similarly. Starting from the definition \eqref{eq: xi-lm-lmlmm-def}, we can write
\beq\label{eq: xi-lmlm-naive}
    \hat\xi_{\ell m\ell'm'}(r,r') &=& \frac{1}{Vv(r)v(r')}\int d\vs'\delta(\vs'+\vr)\delta(\vs'+\vr')Y^*_{\ell m}(\hr)Y^*_{\ell'm'}(\hr') \equiv \frac{1}{V}\int d\vs\,a^*_{\ell m}(\vs;r)a^*_{\ell' m'}(\vs;r'),
\eeq
in terms of the familiar $a_{\ell m}(\vs;r)$ quantities. Since \eqref{eq: xi-lmlm-naive} does not include a weighting $\delta(\vs)$, it is difficult to compute from discrete datasets, since $a_{\ell m}(\vs; r)$ is usually measured at values of $\vs$ corresponding to the locations of galaxies or random particles. An alternative estimator that can be straightforwardly computed from discrete data is obtained by starting from
\beq
    \hat\xi(\vr,\vr') &=& \frac{RNN(\vr,\vr')}{RRR(\vr,\vr')}\equiv \frac{\int d\vs'\, R(\vs')N(\vs'+\vr)N(\vs'+\vr')}{\int d\vs'\,R(\vs')R(\vs'+\vr)R(\vs'+\vr')}.
\eeq
Whilst unconventional (and not strictly minimum-variance), this is unbiased, such that $\av{\hat\xi(\vr,\vr')} = \xi(|\vr-\vr'|)$, which can be easily shown by replacing the variables by their Poisson expectations, \textit{i.e.} $N(\vr)\rightarrow n(\vr)\delta(\vr)$, $R(\vr)\rightarrow n(\vr)$, and averaging over realizations of $\delta$. Analogously to before, we can expand each function as a double spherical harmonic series, giving
\beq
    \sum_{\ell_1m_1\ell_2m_2}\hat\xi_{\ell_1m_1\ell_2m_2}(r_1,r_2)Y_{\ell_1m_1}(\hr_1)Y_{\ell_2m_2}(\hr_2) = \frac{\sum_{L_1M_1L_2M_2}RNN_{L_1M_1L_2M_2}(r_1,r_2)Y_{L_1M_1}(\hr_1)Y_{L_2M_2}(\hr_2)}{\sum_{\ell_1'm_1'\ell_2'm_2'} RRR_{\ell_1'm_1'\ell_2'm_2'}(r_1,r_2)Y_{\ell_1'm_1'}(\hr_1)Y_{\ell_2'm_2'}(\hr_2)}.\nonumber
\eeq
Rearranging this equation and contracting products of spherical harmonics via the Gaunt integral \citep[\S34.3.20]{nist_dlmf} leads to the edge correction equation
\beq\label{eq: xi-lmlm-edge-correction}
    \boxed{\hat\xi_{\ell_1m_1\ell_2m_2}(r_1,r_2) = \sum_{L_1M_1L_2M_2}\left[\bar{\mathsf{M}}^{-1}\right]_{\ell_1m_1\ell_2m_2}^{L_1M_1L_2M_2}(r_1,r_2)\frac{RNN_{L_1M_1L_2M_2}(r_1,r_2)}{RRR_{0000}(r_1,r_2)},}\\\nonumber
    \bar{\mathsf{M}}^{L_1M_1L_2M_2}_{\ell_1m_1\ell_2m_2}(r_1,r_2) = \sum_{\ell_1'm_1'\ell_2'm_2'}\frac{RRR_{\ell_1'm_1'\ell_2'm_2'}(r_1,r_2)}{RRR_{0000}(r_1,r_2)}\mathcal{G}_{\ell_1\ell_1'L_1}^{m_1m_1'-M_1}\mathcal{G}_{\ell_2\ell_2'L_2}^{m_2m_2'-M_2}(-1)^{M_1+M_2}.
\eeq
As in \eqref{eq: NN,RR-def}, the $RNN$ and $NNN$ multiplets may be written
\beq
    RNN_{\ell m\ell'm'}(r,r') &=& \frac{1}{V}\int d\vs'R(\vs')a^*_{\ell m}[N](\vs';r)a^*_{\ell'm'}[N](\vs';r'),\\\nonumber
    RRR_{\ell m\ell'm'}(r,r') &=& \frac{1}{V}\int d\vs'\,R(\vs')a^*_{\ell m}[R](\vs';r)a^*_{\ell'm'}[R](\vs';r').
\eeq
Both are simple to compute alongside the usual $\mathcal{N}_{\ell_1\ell_2\ell_3}$ and $\mathcal{R}_{\ell_1\ell_2\ell_3}$ counts. In detail, the first may be computed by accumulating the sum only if the primary weight is negative, \textit{i.e.} it belongs to the $R(\vs')$ field. Taking the periodic limit offers a simpler expression, as before:
\beq\label{eq: periodic-xi-lmlm}
    \left.RNN_{\ell m\ell'm'}(r,r')\right|_\mathrm{periodic} &=& \frac{\bar{n}^3}{4\pi}\,\delta^\mathrm{K}_{\ell0}\delta^\mathrm{K}_{m0}\delta^\mathrm{K}_{\ell'0}\delta^\mathrm{K}_{m'0}\\\nonumber
    \Rightarrow \left.\hat\xi_{\ell m\ell'm'}(r,r')\right|_\mathrm{periodic} &=& \frac{(4\pi)^2}{\bar{n}^3}RNN_{\ell m\ell'm'}(r,r').
\eeq

In summary, we estimate the connected 4PCF components in the following manner: 
\begin{itemize}
    \item Compute the full 4PCF as usual via \eqref{eq: 4pcf-multiplet-estimator}.
    \item Alongside the $\mathcal{R}_{\ell_1\ell_2\ell_3}$ multiplets, accumulate $RR_{\ell m}(r)$ and $RRR_{\ell m\ell' m'}(r,r')$ contributions.
    \item Alongside the $\mathcal{N}_{\ell_1\ell_2\ell_3}$ multiplets, accumulate $NN_{\ell m}(r)$ and $RNN_{\ell m\ell' m'}(r,r')$ contributions.
    \item Apply edge corrections via \eqref{eq: edge-correction-eq},\,\eqref{eq: xi-lm-edge-correction}\,\&\,\eqref{eq: xi-lmlm-edge-correction} and construct the disconnected piece via \eqref{eq: disconnected-4pcf-estimator}.
    \item Form the connected estimator by subtracting the disconnected piece from the full 4PCF \eqref{eq: connected-4pcf-estimator}.
\end{itemize}

\section{Tests on Lognormal Simulations}\label{sec: lognormals}
Before proceeding to analyze the 4PCF of survey data, we test the estimators of \S\ref{sec: background}\,\&\,\ref{sec: connected-estimator} by applying them to a suite of lognormal simulations at $z = 2$. This has twofold utility: firstly, we may verify that the disconnected estimator is unbiased, \textit{i.e.} that it measures the product of two 2PCFs, projected onto our basis functions; secondly, we may check that the connected contribution is small, as expected at high redshift.

Simulations are generated with \textsc{nbodykit} \citep{2018AJ....156..160H} using the fiducial cosmology of this work, alongside a mean number density $\bar n = 1.5\times 10^{-4}\,h^{-3}\mathrm{Mpc}^3$ and a boxsize $L = 1574\,\Mpch$, matching the number density and volume of the BOSS CMASS sample in the North Galactic Cap (NGC, cf.\,\S\ref{sec: data}). The density fields are generated from a known power spectrum including redshift-space distortions (RSD), using a linear bias of $b = 1.8$. Whilst the high redshift ensures that contributions to the connected 4PCF from the intrinsic galaxy trispectrum are small, we caution that the exponential transform (needed to obtain a discrete density field) gives non-zero $N$-point functions of all order. This further implies that the output 2PCFs will not exactly correspond to those of the input; this mismatch is also sourced by RSD contributions beyond the Kaiser effect. In practice, these effects are found to be small.


For each simulation, we output a set of $\sim$\,$6\times 10^5$ particle positions, which are then combined with sets of uniform random particles. These can be used to compute the $\mathcal{N}_{\ell_1\ell_2\ell_3}$, $RNN_{\ell m\ell'm'}$ and $NN_{\ell m}$ contributions required to form the full and disconnected 4PCF (and thus the connected 4PCF, via \ref{eq: connected-4pcf-estimator}). The computation will be further discussed in \S\ref{subsec: measurements}. Since the mocks have periodic boundary conditions, edge correction of the NPCFs is trivial, and is performed using \eqref{eq: periodic-4pcf},\,\eqref{eq: periodic-xi-lm}\,\&\,\eqref{eq: periodic-xi-lmlm}. All calculations are implemented in the \textsc{encore} code \citep{npcf_algo}, and use $N_r = 10$ radial bins per dimension (with $r_{\rm min} = 20\,\Mpch$ and $r_{\rm max} = 160\,\Mpch$) including all multiplets up to $\ell_{\rm max} = 4$, giving a total of $N_\zeta = 5040$ 4PCF bins (given the above restrictions on $\ell_i$ and $r_i$). Computation of the full and disconnected 4PCFs from 1\,000 lognormal simulations required $\sim$\,$13$k CPU-hours overall.  

To validate the disconnected 4PCF estimator, we require a theory model. As noted above, the disconnected piece is simply the product of two 2PCFs, \textit{i.e.} $\zeta^{(\rm disc)}(\vr_1,\vr_2,\vr_3) = \xi(\vr_1)\xi(\vr_2-\vr_3) + \text{2 perms}$, projected onto the isotropic basis functions of \eqref{eq: N=4-basis}. Following a short calculation described in Appendix \ref{appen: gaussian-deriv}, we find the following model for the disconnected contribution (or equivalently, the full 4PCF in the Gaussian limit):
\begin{empheq}[box=\fbox]{align}\label{eq: gaussian-4pcf-model}
    &\zeta^\mathrm{G}_{\ell_1\ell_2\ell_3}(r_1,r_2,r_3) = (4\pi)^{3/2}\sqrt{(2\ell_1+1)(2\ell_2+1)(2\ell_3+1)}\tjo{\ell_1}{\ell_2}{\ell_3}\\\nonumber
    &\,\times\,\left\{\frac{i^{\ell_2-\ell_3}}{(2\ell_1+1)^2}\,\xi_{\ell_1}(r_1)f^{\ell_1}_{\ell_2\ell_3}(r_2,r_3)+\frac{i^{\ell_1-\ell_3}}{(2\ell_2+1)^2}\,\xi_{\ell_2}(r_2)f^{\ell_2}_{\ell_1\ell_3}(r_1,r_3)+\frac{i^{\ell_1-\ell_2}}{(2\ell_3+1)^2}\,\xi_{\ell_3}(r_3)f^{\ell_3}_{\ell_1\ell_2}(r_1,r_2)\right\},
\end{empheq}
\eqref{eq: gaussian-4pcf-full-appen}, defining
\beq
    f^{L}_{\ell\ell'}(r,r') = \int\frac{k^2dk}{2\pi^2}P_L(k)j_{\ell}(kr)j_{\ell'}(kr'),
\eeq
where $j_\ell(x)$ is a spherical Bessel function of order $\ell$ and $P_L$ ($\xi_L$) is the $L$-th Legendre multipole of the non-linear power spectrum (2PCF), assuming a fixed line-of-sight.\footnote{We note that the 2PCF multipoles inherit a factor of $i^\ell$ relative to the power spectrum multipoles due to the definition of \eqref{eq: xi-ell,f-ell-binned}.} Importantly, the \textit{isotropic} 4PCF contains contributions from the \textit{anisotropic} 2PCF multipoles; this is analogous to the anisotropic contributions that enter the covariance matrix of the isotropic 2PCF \citep[e.g.,][]{2016MNRAS.457.1577G}. Assuming that only even 2PCF multipoles are non-zero and noting that the 3-$j$ symbol requires $(-1)^{\ell_1+\ell_2+\ell_3}=1$, we find that the Gaussian 4PCF is explicitly real. For comparison to data, the theory model must additionally be binned in radius. This procedure is described in Appendix \ref{appen: gaussian-deriv} and simply leads to the replacement of $\xi_\ell$ and $f_{\ell_1\ell_2}^L$ in \eqref{eq: gaussian-4pcf-model} by their bin-averaged forms \eqref{eq: xi-ell,f-ell-binned}, involving (analytic) bin-averaged spherical Bessel functions. 


\begin{figure}
    \centering
    \includegraphics[width=0.8\textwidth]{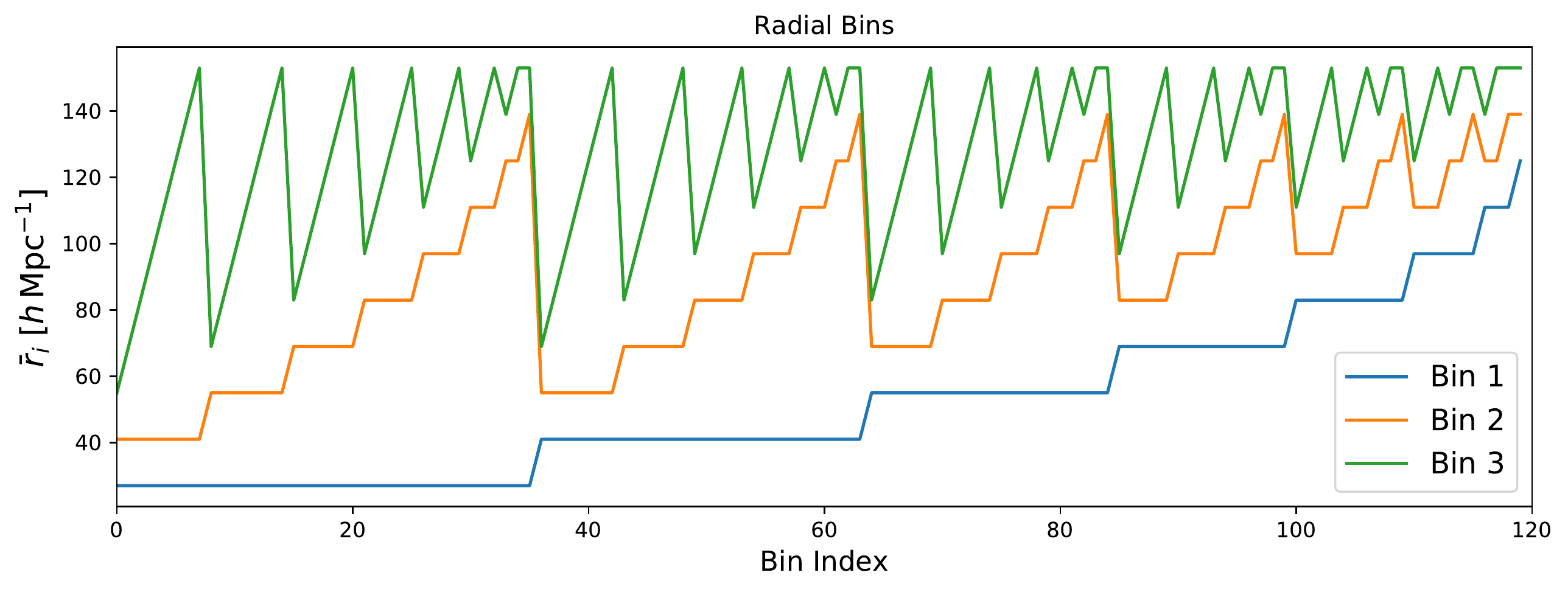}\\
    \includegraphics[width=0.8\textwidth]{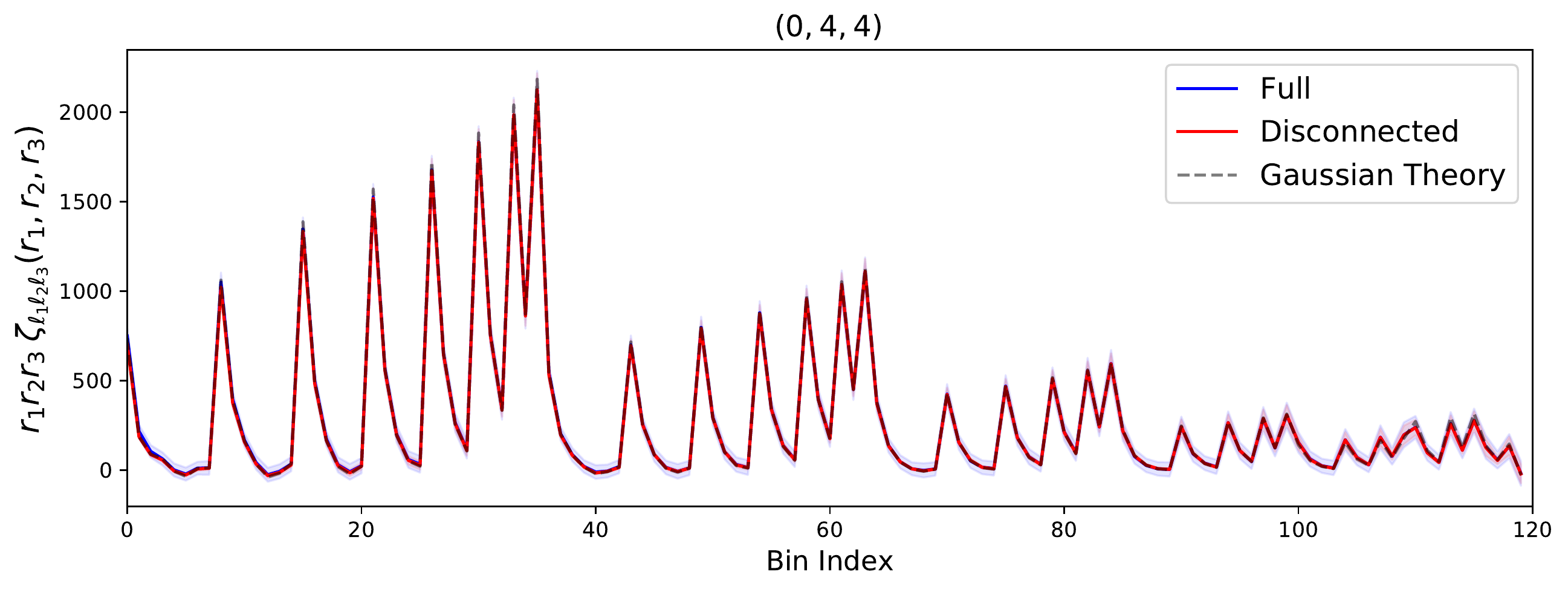}\\
    \includegraphics[width=0.8\textwidth]{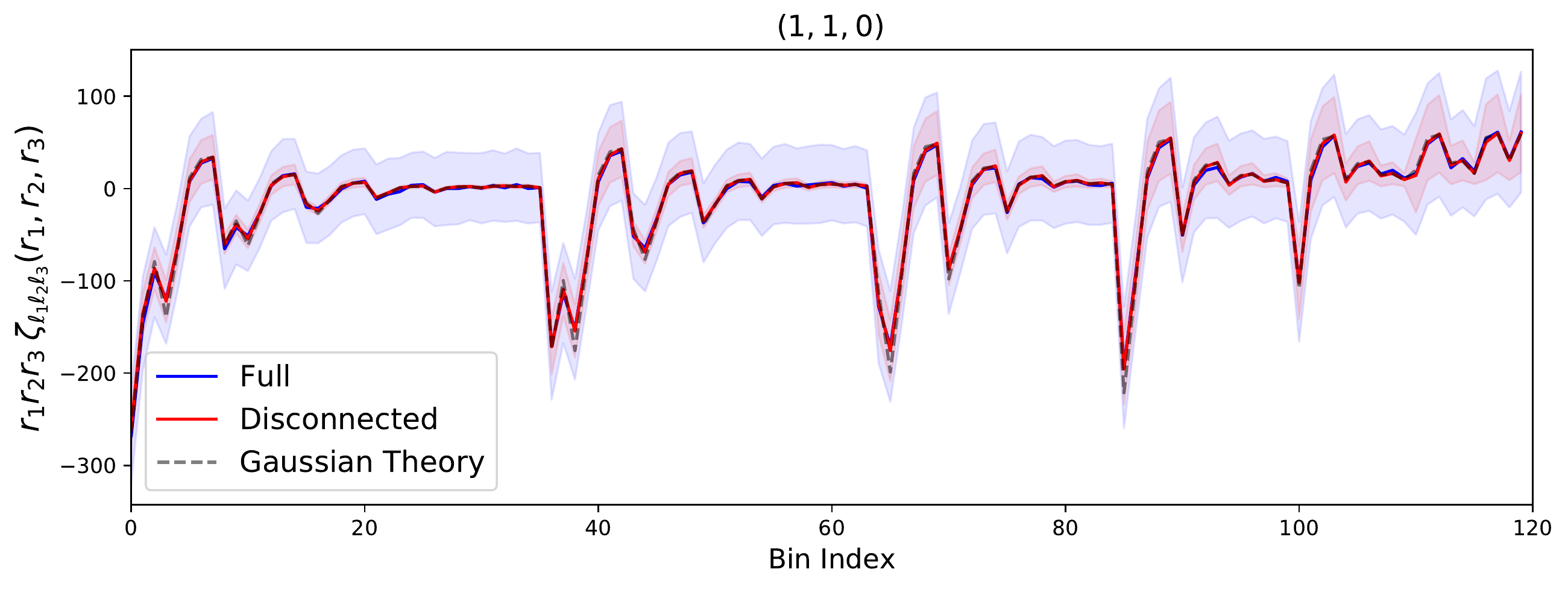}\\
    \includegraphics[width=0.8\textwidth]{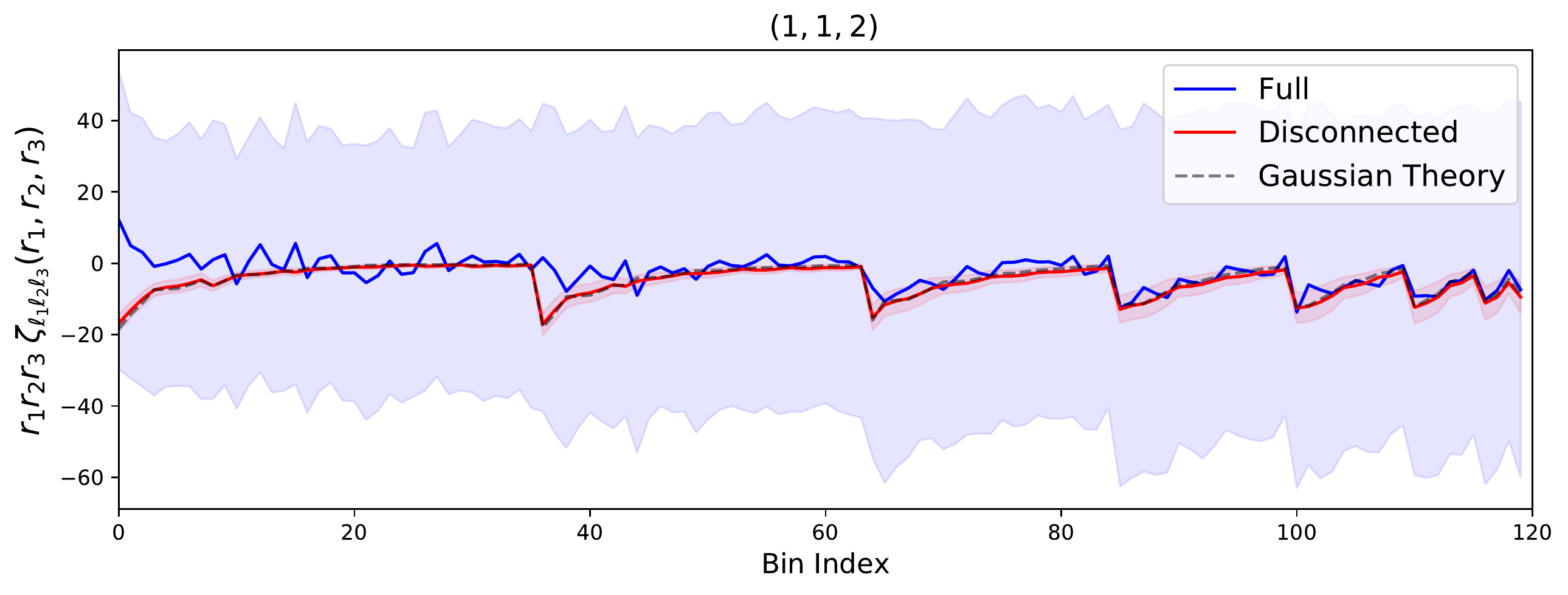}\\
    \caption{\small Full and disconnected 4PCF multiplets estimated from a suite of 1\,000 lognormal mocks at $z = 2$. The first panel shows the values of the linearly-averaged radial bin-centers, with the horizontal axis giving the index of the bins, collapsed into one dimension. Bins are ordered in increasing radius, and satisfy $r_1<r_2<r_3$. In the succeeding panels, we display the 4PCF of three representative multiplets (\textit{i.e.} $\zeta_{\ell_1\ell_2\ell_3}$ for different choices of $\ell_1,\ell_2,\ell_3$), with the multiplets indicated by the title. Measurements using the full 4PCF estimator of \S\ref{sec: background} are shown in solid blue lines, with those using the disconnected estimator of \S\ref{sec: connected-estimator} shown in solid red lines. In both cases the mean values are shown as lines with the shaded regions giving the $1\sigma$ errors. We further show the Gaussian theory model in dashed gray lines. The disconnected 4PCF closely matches the theory model, and furthermore, the connected 4PCF (equal to the difference between full and disconnected contributions) is small.}
    \label{fig: 4pcf-lognormal-mult}
\end{figure}


In Fig.\,\ref{fig: 4pcf-lognormal-mult}, we compare the measured 4PCF multiplets from lognormal simulations with those of the analytic model \eqref{eq: gaussian-4pcf-model}, computed using the input power spectrum multipoles $P_L(k)$. We find good agreement between the disconnected measurements and the Gaussian theory model, affording us confidence that the disconnected 4PCF estimator works as expected. As noted in Appendix \ref{appen: gaussian-deriv}, if the 2PCF is isotropic, only multiplets of the form $\{0,\ell,\ell\}$, $\{\ell,0,\ell\}$ or $\{\ell,\ell,0\}$ are non-zero; the clear non-zero $\zeta_{112}$ contributions shown in Fig.\,\ref{fig: 4pcf-lognormal-mult} highlight the importance of incorporating 2PCF anisotropies into the modelling. Furthermore, the connected contributions (equal to the difference between the full and disconnected estimator) are found to be very small. Again, this validates our procedure, since we are working at high redshift, whereupon the trispectrum contributions should be strongly suppressed. Motivated by these conclusions, we will henceforth display results only from the \textit{connected} 4PCF estimator.

\section{Analysis Techniques}\label{sec: techniques}
Given the estimators of \S\ref{sec: background}\,\&\,\ref{sec: connected-estimator} and their verification in \S\ref{sec: lognormals}, we are now in a position to consider how best to analyze the 4PCF of real surveys. 
Classically, cosmological data are analyzed by way of the $\chi^2$ statistic, defined as
\beq\label{eq: classical-chi2}
    \chi^2 \equiv \left(\hat\zeta^{(\rm c)}-\zeta^{(\rm c)}_{\rm model}\right)^T\mathsf{C}^{-1}\left(\hat\zeta^{(\rm c)}-\zeta^{(\rm c)}_{\rm model}\right),
\eeq
treating the connected 4PCF measurements as an $N_\zeta$-dimensional vector with covariance $\mathsf{C}$, implicitly summing over multiplets and bins. Here, our null hypothesis is that there is no connected 4PCF, hence $\zeta_{\rm model}^{(\rm c)} = \vec 0$. It is common to use the sample covariance $\hat{\mathsf{S}}$ as a proxy for the true covariance:
\beq\label{eq: sample-cov}
    \hat{\mathsf{S}} = \frac{1}{N_\mathrm{mocks}-1}\sum_{i=1}^{N_{\rm mocks}}\left(\zeta^{(\rm c)}_i-\bar\zeta^{(\rm c)}\right)\left(\zeta^{(\rm c)}_i-\bar\zeta^{(\rm c)}\right)^T,
\eeq
where $\{\zeta^{(\rm c)}_i\}$ are the 4PCF measurements from a suite of $N_{\rm mocks}$ mocks with mean $\bar\zeta^{(\rm c)}$. This gives the $\chi^2$ estimate
\beq\label{eq: classical-sample-chi2}
    \hat\chi^2 = \hat\zeta^{(\rm c),T}\hat{\mathsf{S}}^{-1}\hat\zeta^{(\rm c)}.
\eeq

This approach bears a number of problems: (a) the sample covariance is invertible only if $N_{\rm mocks}>N_{\zeta}$, which is difficult to achieve given the high dimensionality of the 4PCF, (b) noise in the sample covariance matrix gives an additional source of noise in $\chi^2$, changing its probability distribution \citep{2016MNRAS.456L.132S}, and (c) an underlying assumption is that the 4PCF is Gaussian-distributed, which is questionable \citep[e.g.,][]{2003ApJ...592..664P,2012PhRvD..86f3009S,2019PhRvD.100l3511M,2019MNRAS.485.2956H}. To ameliorate (a) and (b), one can replace $\hat{\mathsf{S}}$ with some \textit{theoretical} covariance matrix (cf.\,\S\ref{subsec: covariance}), which is straightforwardly invertible.\footnote{One may also use an improved estimator for the sample covariance, such as shrinkage estimators \citep{2017MNRAS.466L..83J}, which do not require $N_{\rm mocks}>N_{\zeta}$. Unlike the procedure outlined in \S\ref{subsec: compression}, this does not guarantee that the 4PCF detection is unbiased.} However, with realistic survey geometries, the theoretical covariance is known to differ significantly from the measured covariance \citep{npcf_cov}. In this case, the expectation of \eqref{eq: classical-chi2} is not equal to $N_\zeta$ even under the null hypothesis of $\av{\hat\zeta^{(\rm c)}} = \vec 0$, thus it may lead to a spurious detection of non-Gaussianity. In \citep{npcf_cov} (and to an extent in \citep{2016MNRAS.462.2681O,2020MNRAS.491.3290P,2020PhRvD.102l3521W}), it was suggested that altering the survey volume $V$ and number density $\bar n$ could partially absorb the effects of non-Gaussianity and window function non-uniformity; whilst this results in visually similar covariances, it can lead to spurious detections of non-Gaussianity, as discussed in Appendix \ref{appen: rescaled-theory-cov}.

\subsection{Data Compression}\label{subsec: compression}
A convenient manner in which to deal with the curse of dimensionality is to compress the 4PCF data-vector $\zeta^{(\rm c)}$. Many compression schemes exist in the literature \citep[e.g.,][]{2000MNRAS.317..965H,2018MNRAS.476L..60A,2021PhRvD.103d3508P,2019MNRAS.484L..29G}, all of which aim to reduce the dimension of the observable whilst retaining information contained within the Fisher matrix or log-likelihood. Since our goal is to simply make a measurement of non-Gaussianity, quantified by a $\chi^2$-like test under the null hypothesis, we will adopt a procedure similar to \citep{2000ApJ...544..597S,2013MNRAS.432.1928T}, expanding the signal in a basis of approximate eigenmodes, then selecting those with maximal signal-to-noise. A key part of this is a \textit{theoretical} covariance matrix; here, we use that of \citep{npcf_cov}, which will be discussed further in \S\ref{subsec: covariance}. 

Explicitly, we first perform an eigendecomposition of the theory covariance matrix via $\mathsf{C}_{\rm theory} = \mathsf{U}\mathsf{D}\mathsf{U}^T$, where $\mathsf{U}$ is the unitary matrix of eigenvectors and $\mathsf{D}$ is a diagonal matrix of eigenvalues. Since the theory covariance is of full rank, we expect all eigenvalues to be non-zero. The 4PCF is then projected onto the eigenvectors via
\beq
    v \equiv \mathsf{U}^T\zeta^{(\rm c)}.
\eeq
Finally, we perform dimensionality reduction by restricting to the first $N_{\rm eig}$ eigenvectors, as ordered by signal-to-noise:
\beq
    \left(\frac{S}{N}\right)_i = \frac{|\bar{v}_i|}{D_i},
\eeq
where $\bar{v}$ is the mean 4PCF from a suite of realistic mocks (which we assume to be non-zero). Assuming $\mathsf{C}_{\rm theory}$ to be close to the true covariance, the compression scheme picks out the basis vectors that contribute most to the detection of non-zero $\chi^2$ in the data. Had our goal been to derive parameter constraints from the 4PCF, this decomposition would not be optimal, since the components contributing most to the signal-to-noise are not necessarily those with maximal information about the parameters of interest. In such a case, a Fisher- or likelihood-based compression scheme such as \citep{2000MNRAS.317..965H} or \citep{2021PhRvD.103d3508P} would be preferred.

Following the compression, analysis proceeds in the low-dimensional subspace. In particular, we can define a new $\chi^2$ variable by compressing the mock 4PCF measurements and defining a sample covariance in the standard fashion (cf.\,\ref{eq: sample-cov}):
\beq
    \hat{\mathsf{S}}_{v} = \frac{1}{N_{\rm mocks}}\sum_{i=1}^{N_{\rm mocks}} \left(v_i-\bar{v}\right)\left(v_i-\bar{v}\right)^T \equiv \mathsf{U}^T\hat{\mathsf{S}}\mathsf{U}, 
\eeq
assuming $\mathsf{U}$ to be of dimension $N_{\zeta} \times N_{\rm eig}$, \textit{i.e.} dropping all but the highest signal-to-noise eigenmodes. If $N_{\rm mocks}>N_{\rm eig}$ this is invertible, meaning that a mock-based analysis is possible. The compressed $\chi^2$ is given by (cf.\,\ref{eq: classical-sample-chi2})
\beq\label{eq: compressed-chi2}
    \hat{\chi}^2_v = \hat v^T\hat{\mathsf{S}}_v^{-1}\hat v \equiv \hat\zeta^{(\rm c),T}\mathsf{U}\left[\mathsf{U}^T\hat{\mathsf{S}}\mathsf{U}\right]^{-1}\mathsf{U}^T\hat\zeta^{(\rm c)}.
\eeq
Since we apply the same compression to the data, the covariance, and the model, the compression cannot generate any spurious 4PCF signals, \textit{i.e.} it is unbiased for any choice of projection matrix $\mathsf{U}$. 
In fact, the only effect of data compression is to lessen the significance of any detection. To see this, consider the case when $N_{\rm mocks}$ is large, such that $\hat{\mathsf{S}} \approx \mathsf{C}$. The signal-to-noise of some true 4PCF $\bar \zeta^{(\rm c)}$ is given by
\beq
    \left(\frac{S}{N}\right)^2 = \bar\zeta^{(\rm c)}\mathsf{C}^{-1}\bar\zeta^{(\rm c)}, \qquad \left(\frac{S}{N}\right)^2_v = \bar\zeta^{(\rm c),T}\mathsf{U}\left[\mathsf{U}^T\mathsf{C}\mathsf{U}\right]^{-1}\mathsf{U}^T\bar\zeta^{(\rm c)},
\eeq
pre- and post-compression respectively. Using the properties of projection matrices, we find that
\beq
    \left(\frac{S}{N}\right)^2 \geq \left(\frac{S}{N}\right)^2_v
\eeq
for arbitrary $\bar\zeta^{(\rm c)}$, with equality only if $N_{\rm eig} = N_\zeta$, \textit{i.e.} without projection.\footnote{This can be straightforwardly shown by diagonalizing the covariances of $\zeta$ and $v$.} Thus, the compression can lead only to a reduced signal-to-noise; further, this is minimized by our choice of projection matrix $\mathsf{U}$. By the Eckart-Young theorem \citep{eckart-young}, our compression is optimal (in terms of signal-to-noise) if $\mathsf{C}_{\rm theory} = \mathsf{C}$. 
In practice, increasing $N_{\rm eig}$ might \textit{not} lead to an increased detection significance, due to the broadening of the $\chi^2$ posterior when using a finite number of mocks, as we further discuss below.


\subsection{Probability Distribution}
If the data are Gaussian-distributed and the covariance is known precisely (\textit{i.e.} in the limit $N_{\rm mocks}\to\infty$), the compressed $\chi^2$ statistic of \eqref{eq: compressed-chi2} follows the $\chi^2$ distribution with $N_{\rm eig}$ degrees of freedom. In the case of a finite number of mocks, we must account for the effects of noise in the sample covariance, which changes the distribution of \eqref{eq: compressed-chi2}. A commonly-used approach is to rescale \eqref{eq: compressed-chi2} by the \textit{Hartlap} factor, \textit{i.e.}
\beq\label{eq: hartlap-chi2}
    \hat\chi^2_v \to \frac{N_{\rm mocks}-N_{\rm eig}-2}{N_{\rm mocks}-1}\hat\chi^2_v \equiv \hat{H}^2_v
\eeq
\citep{2007A&A...464..399H,1933PCPS...29..260W}, which debiases the inverse sample covariance. In this case, $\hat{H}^2_p$ is then analyzed using the $\chi^2$ distribution with $N_{\rm eig}$ degrees of freedom.

As shown in \citep{2016MNRAS.456L.132S},\footnote{See also \citep{2013MNRAS.432.1928T} for an approximate treatment in the Gaussian limit.} this is not a consistent manner in which to treat noise in the sample covariance. Instead, the data follow the $t$-distribution, with the PDF
\beq\label{eq: T2-pdf}
    f_T(T^2;n,p) = \frac{\Gamma\left([n+1]/2\right)}{\Gamma(p/2)\Gamma\left([n-p+1]/2\right)}\frac{n^{-p/2}(T^2)^{p/2-1}}{(T^2/n+1)^{(n+1)/2}}
\eeq
\citep{2016MNRAS.456L.132S}, denoting the sample statistic of \eqref{eq: compressed-chi2} by $T^2$ and writing $n = N_{\rm mocks}-1$ and $p=N_{\rm eig}$ for Gamma function $\Gamma$. At finite $N_{\rm mocks}$, $f_T$ has a greater width than the $\chi^2$ distribution and will be used to avoid false signal detections.

\subsection{Likelihood Non-Gaussianity}\label{subsec: non-gaussian-likelihood}
An underlying assumption in the above sections is that the likelihood for the 4PCF is Gaussian; \textit{i.e.} that $\hat\zeta$ is a draw from some multivariate normal distribution with covariance $\mathsf{C}$. However, higher-order statistics such as the NPCFs are known to have increasingly non-Gaussian distributions as $N$ increases \citep[e.g.,][]{2003ApJ...592..664P,2012PhRvD..86f3009S,2019PhRvD.100l3511M,2019MNRAS.485.2956H}, thus performing a $\chi^2$ analysis (or a variant such as \ref{eq: T2-pdf}) may significantly over- (or under-)estimate the detection significance. In general, empirical determination of likelihood non-Gaussianity is difficult (though see \citep{2019MNRAS.485.2956H} for an interesting approach) and beyond the scope of this work. Since the bulk of our analysis is performed on compressed statistics with $\mathcal{O}(100)$ elements, we expect the Central Limit Theorem to significantly reduce any non-Gaussianity present. A simple check of this reduction is to take a set of simulations and split them into two pieces, one of which is used to determine the sample mean and covariance, and the other of which is used as mock data. One may then compare the empirical distribution of $\hat\chi^2_v$ (including the sample mean) to the expected distribution \eqref{eq: T2-pdf}. A visual comparison suggests that the degree of non-Gaussianity is not of particular importance; that said we encourage the reader to take the exact detection significances quoted below with a grain of salt, and hope to examine these effects in more detail in future work.


\section{Practical Application}\label{sec: data}

Below, we consider the practicalities of computing and analyzing the 4PCF from observational data, including discussion of the theoretical covariance matrix.

\subsection{Data and Simulations}\label{subsec: data}
Our dataset comprises galaxies from the twelfth data release (DR12) \citep{2015ApJS..219...12A} of the Baryon Oscillation Spectroscopic Survey (BOSS), part of SDSS-III \citep{2011AJ....142...72E,2013AJ....145...10D}.\footnote{The BOSS data is publicly available at \href{https://data.sdss.org/sas/dr12/boss/lss/}{data.sdss.org/sas/dr12/boss/lss/}.} In particular, we use the CMASS region, containing a total of 587\,071 (216\,041) galaxies in the redshift range $z_\mathrm{min} = 0.43$ to $z_\mathrm{max} = 0.7\}$ from the North (South) Galactic Cap (hereafter denoted `NGC' and `SGC').\footnote{Technically, we use the `CMASSLOWZTOT' catalogs, with the CMASS redshift cuts. This is to ensure consistency between data and simulations.} Here, we use both the galaxy data (converting angles and distances into Cartesian coordinates using our fiducial cosmology) and a set of unclustered randoms of $\sim$\,$50\times$ greater number density. Galaxies are assigned weights as in \citep{2017MNRAS.466.2242B}:
\beq
    w_\mathrm{tot} = (w_\mathrm{rf}+w_\mathrm{fc}-1)w_\mathrm{sys}w_\mathrm{fkp}
\eeq
where $w_\mathrm{rf}$, $w_\mathrm{fc}$ and $w_\mathrm{sys}$ correspond to redshift-failure, fiber-collision and systematic weights, with $w_\mathrm{fkp} = [1+n(z)P_0]^{-1}$ being the FKP weight \citep{1994ApJ...426...23F} for background number density $n(z)$ and $P_0 = 10^4\,h^{-3}\mathrm{Mpc}^3$. The FKP weight ensures optimal spectrum extraction on small scales.\footnote{For a discussion of optimal analyses on more general scales, see \citep{2021PhRvD.103j3504P}.} Similar considerations apply for the random weights, except that the BOSS randoms do not contain any completeness or systematic weights.

To generate sample covariance matrices for our 4PCF statistic (required in \S\ref{sec: techniques}), we use a suite of $1\,000$ \textsc{MultiDark-Patchy} (hereafter `Patchy') simulations \citep{2016MNRAS.456.4156K,2016MNRAS.460.1173R}, with the same survey geometry and radial selection function as the BOSS sample, both for the NGC and SGC regions. Weights are assigned via
\beq
    w_\mathrm{tot} = w_\mathrm{veto}w_\mathrm{fc}w_\mathrm{fkp}
\eeq
where $w_\mathrm{veto}$ is the veto flag (either zero or one), arising, from, for example, bright star masks. The Patchy simulations additionally have a separate random catalog.

\subsection{4PCF Computation}\label{subsec: measurements}
As in \S\ref{sec: lognormals}, we compute the 4PCF multiplets, $\zeta_{\ell_1\ell_2\ell_3}^{abc}$, using the \textsc{encore} code \citep{npcf_algo}.\footnote{\href{https://github.com/oliverphilcox/encore}{github.com/oliverphilcox/encore}}
As in \S\ref{sec: lognormals}, we use $N_r = 10$ radial bins per dimension with $r_\mathrm{min} = 20\,\Mpch$, $r_\mathrm{max} = 160\,\Mpch$. Given the additional restriction $r_1<r_2<r_3$, in order to avoid zero-separation bins, we have a total of 120 radial components per multiplet. To accurately capture the dominant signal contributions, we compute all $\{\ell_1,\ell_2,\ell_3\}$ multiplets up to $\ell_\mathrm{max} = 5$, but use the $\ell=5$ elements only for edge correction.  In total, we estimate $8\,280$ 4PCF components, of which $5\,040$ are used in the analysis of \S\ref{sec: results}. Whilst increasing the number of bins ensures that any fine features (such as BAO) are fully resolved, 
this incurs a significant computational penalty both in dimensionality of the output statistic (which is bounded from above by $N_r^3[\ell_\mathrm{max}+1]^{3}$) and in computation time (which scales as $N_{\rm g}N_r^3(\ell_\mathrm{max}+1)^5$ for $N_{\rm g}$ galaxies).

For computational ease, we split the random catalogs into 32 chunks, each containing $1.5\times$ more randoms than galaxies, and analyze each corresponding `data-minus-random' catalog separately, using \eqref{eq: data-minus-randoms}.\footnote{\citep{2015MNRAS.454.4142S} found this to produce an optimal noise-to-computation-time ratio for the 3PCF. Since the computation time of the 4PCF is usually found to scale as $\mathcal{O}(N_{\rm g})$ rather than $\mathcal{O}(N_{\rm g}^2)$, the total work is broadly independent of the partition size.} Notably, we store the $a^i_{\ell m}(\vs)$ counts \eqref{eq: discrete-4pcf-counts} from each `data-data' particle pair after the first iteration to avoid unnecessary re-computation. All simulations are analyzed on a 16-core Intel processor (of various generations), embarrassingly parallelized using \texttt{OpenMP}. We simultaneously compute both the full 4PCF counts \eqref{eq: edge-correction-eq} and the two- and three-particle counts required to compute the disconnected piece \eqref{eq: xi-lm-edge-correction}\,\&\,\eqref{eq: xi-lmlm-edge-correction}, which are combined straightforwardly in post-processing using \textsc{Python}. In total, computation of the connected 4PCF requires $\sim$\,$40$\,($6$) CPU-hours per NGC (SGC) simulation, or a total of $\sim$\,$50$k\,CPU-hours for the complete set of data and 1\,000 Patchy mocks. The inclusion of the disconnected component is found to increase the runtime by $\sim 20\%$. Whilst not insignificant, the runtime is comparable to the computational cost of the 2PCF analysis in \citep{2018MNRAS.477.1153V}. 

\subsection{Analytic Covariance Matrices}\label{subsec: covariance}
To facilitate the data compression techniques discussed in \S\ref{subsec: compression}, we require a smooth approximation to the connected 4PCF covariance matrix. For this purpose, we will use that derived in \citep{npcf_cov}, under the assumptions of isotropy (\textit{i.e.} no redshift-space distortions), Gaussianity (\textit{i.e.} no contributions from NPCFs with $N>2$) and a uniform survey geometry. Whilst these assumptions lead to a covariance that does not precisely match the sample covariance of a realistic survey (as found in \citep{npcf_cov} for the \textsc{MultiDark-Patchy} simulations, and demonstrated here in Fig.\,\ref{fig: cov-comparison}), they do offer an analytically tractable theory. We recall that our compression scheme does \textit{not} require the theoretical covariance to be unbiased; rather it should be smooth and 
have eigenvectors close to those of the true covariance.

Before projection onto the angular basis, the covariance matrix of the full 4PCF (including both connected and disconnected contributions) is defined by:
\beq\label{eq: basic-cov-def}
    \mathrm{Cov}(\vr_1,\vr_2,\vr_3;\vr_1',\vr_2',\vr_3') = \av{\hat\zeta(\vr_1,\vr_2,\vr_3)\hat\zeta^*(\vr_1',\vr_2',\vr_3')} - \av{\hat\zeta(\vr_1,\vr_2,\vr_3)}\av{\hat\zeta^*(\vr_1',\vr_2',\vr_3')}.
\eeq
Inserting \eqref{eq: 4pcf-basic-estimator}, this involves the expectation of eight density fields. As noted in \citep{npcf_cov}, the Gaussian covariance contains two contributions: `fully coupled' (including correlations only between density fields in different 4PCFs, e.g., between $\delta(\vs+\vr_1)$ and $\delta(\vs'+\vr_1')$), and `partially coupled' (including one correlation within each 4PCF, e.g., between $\delta(\vs+\vr_1)$ and $\delta(\vs+\vr_2)$). Respectively, these give the covariance contributions
\beq\label{eq: basis-cov-split}
    \mathrm{Cov}^{(fc)}\left(\vr_1,\vr_2,\vr_3,\vr_1',\vr_2',\vr_3'\right) &=& \frac{1}{V}\int d\vs\,\xi(\vs+\vr_0-\vr_0')\xi(\vs+\vr_1-\vr_1')\xi(\vs+\vr_2-\vr_2')\xi(\vs+\vr_3-\vr_3')+\text{23 perms.}\nonumber\\
    \mathrm{Cov}^{(pc)}\left(\vr_1,\vr_2,\vr_3,\vr_1',\vr_2',\vr_3'\right) &=& \xi(\vr_1-\vr_0)\xi(\vr_1'-\vr_0')\,\frac{1}{V}\int d\vs\,\xi(\vs+\vr_2-\vr_2')\xi(\vs+\vr_3-\vr_3') + \text{71 perms.},
\eeq
via Wick's theorem. To simplify the permutation structure of \eqref{eq: basis-cov-split}, we have introduced dummy variables $\vr_0$ and $\vr_0'$; these can be later set to zero. Here, we require the covariance of the \textit{connected} 4PCF estimator \eqref{eq: connected-4pcf-estimator}, which requires only the `fully-coupled' terms of \eqref{eq: basis-cov-split} \citep{npcf_cov}. All partially-coupled terms are cancelled by subtraction of the disconnected piece \eqref{eq: connected-4pcf-estimator}, which significantly reduces the estimator covariance.\footnote{Strictly this is true only in the limit that the volume $V\gg r_c^3$ where $r_c$ is the typical correlation scale. Assuming $r_c$\,$\sim$\,$100\,\Mpch$ (the BAO scale), this ratio is $\ll 1\%$ in most cases and thus may be safely neglected.} An explicit demonstration of this is given in Appendix \ref{appen: fully-coupled-covariance}.

The above argument extends naturally to the 4PCF multiplets. In this case, the covariance becomes
\beq\label{eq: multiplet-covariance}
    \mathrm{Cov}_{\ell_1^{}\ell_2^{}\ell_3^{};\,\ell_1'\ell_2'\ell_3'}(r_1,r_2,r_3;r_1',r_2',r_3') &=& \int \left[\prod_{i=1}^3 d\hr_i\,d\hr'_i\right]\,\mathrm{Cov}\left(\vr_1,\vr_2,\vr_3,\vr_1',\vr_2',\vr_3'\right)\\\nonumber
    &&\,\times\,\P_{\ell_1\ell_2\ell_3}^*(\hr_1,\hr_2,\hr_3)\P_{\ell_1\ell_2\ell_3}^*(\hr_1',\hr_2',\hr_3');
\eeq
the simplified forms for this are given explicitly in \cite{npcf_cov}. It is straightforward to include radial binning in the covariance by replacing any spherical Bessel functions in \eqref{eq: multiplet-covariance} with their bin-averaged equivalents as in \eqref{eq: bin-averaged-sBF}. Both basis projection and radial binning are included in the covariances used in \S\ref{sec: results}.

To implement the Gaussian covariance, we require a model for the (survey-geometry-corrected) galaxy power spectrum. Here, we first compute the power spectrum multipoles, $P_\ell(k)$ (for $\ell\in\{0,2,4\}$), of the Patchy simulations using \textsc{nbodykit} \citep{2018AJ....156..160H}, adopting a $k$-space binning of $k_\mathrm{min} = 0.005\hMpc$, $k_\mathrm{max} = 0.25\hMpc$ and $\Delta k = 0.005\hMpc$. Since this work uses the CMASS sample, rather than the `z1' and `z3' split found in BOSS power spectrum analyses \citep[e.g.,][]{2017MNRAS.466.2242B,2020JCAP...05..042I}, we recompute the window function multipoles, $RR_{\ell}(r)$, by performing finely binned angular and radial pair-counts of the random particle catalog using \textsc{corrfunc} \citep{2020MNRAS.491.3022S}, downsampling the randoms by a factor of five for tractability. As in \citep{2017MNRAS.464.3121W,2017MNRAS.466.2242B}, these facilitate comparison of the unwindowed power spectrum model and the window-convolved data. Data are fitted using a one-loop Effective Field Theory of Large Scale Structure model using the \textsc{class-pt} code \citep{2020JCAP...04..042C} and the bias parametrization of \citep{2020JCAP...05..042I}.\footnote{See \href{https://github.com/Michalychforever/CLASS-PT}{github.com/Michalychforever/CLASS-PT} with likelihoods available at \href{https://github.com/Michalychforever/lss_montepython}{github.com/Michalychforever/lss\_montepython}.} This was recently shown to have exquisite accuracy on huge volume simulations \citep{2020PhRvD.102l3541N}, and we follow the treatment and bias parametrization of \citep{2020JCAP...05..042I}. The unwindowed best-fit power spectrum is then used as an input to the analytic covariance. For simplicity, we compute only a single theory model using the NGC data, since the two hemispheres are found to have similar bias parameters \citep{2020JCAP...05..042I}. To define the volume and shot-noise of each region we use the approach of \citep[Appendix A]{2020PhRvD.102l3521W} involving the random particle catalog, as described in \citep{npcf_cov}.

\section{Results}\label{sec: results}

\begin{figure}
    \centering
    \includegraphics[width=0.95\textwidth]{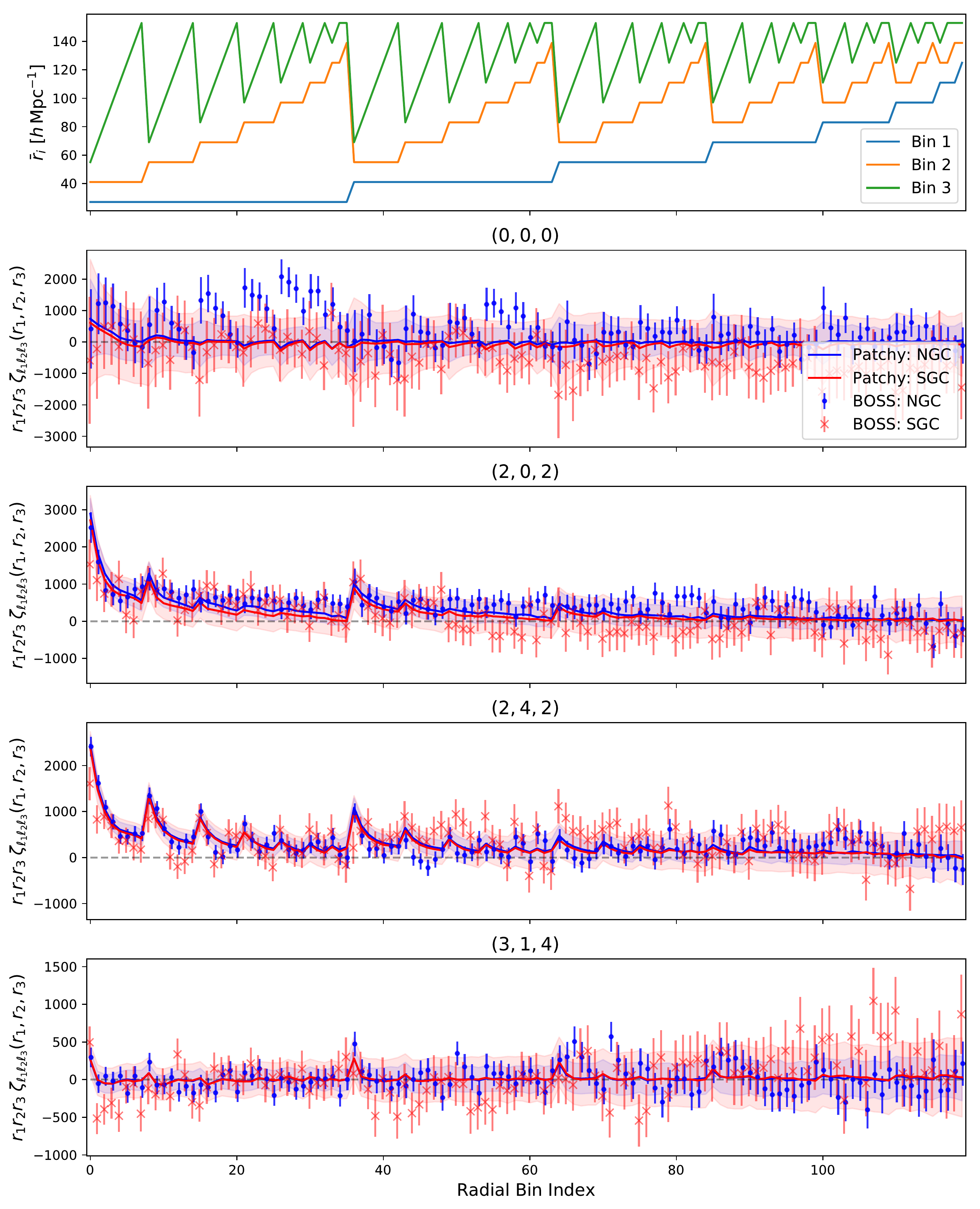}
    \caption{Connected 4PCF measurements from the BOSS CMASS dataset, alongside a suite of 1\,000 Patchy mocks. Results from the NGC (SGC) region are shown in blue (red), with the lines and shaded regions showing the mean and $1\sigma$ deviations from Patchy respectively. As in Fig.\,\ref{fig: 4pcf-lognormal-mult}, the first panel shows the radial bins corresponding to each one-dimensional bin index, and the lower panels give the results for a subset of the 42 measured multiplets, $\zeta_{\ell_1\ell_2\ell_3}$ (rescaled by $r_1r_2r_3$), with the values of $\ell_1,\ell_2,\ell_3$ given in the title. Though the measurements are both highly correlated and noisy, there is a clear non-zero signal observed, particularly in bins where the three side lengths are small. Any such detection is evidence of (gravitationally-induced) non-Gaussianity in the dataset.}
    \label{fig: boss-4pcfs}
\end{figure}

In Fig.\,\ref{fig: boss-4pcfs}, we show a selection of the 42 connected 4PCF multiplets measured from the BOSS dataset and the Patchy simulations. Considering first Patchy, we find that, whilst the measurements are noisy, those from the mean-of-the-mocks show strong evidence for a non-zero connected 4PCF, \textit{i.e.} for gravitational non-Gaussianity. This is enhanced on scales with small tetrahedron side lengths (\textit{i.e.} small $\bar r_i$ in the figure); since structure forms first on small scales, this is as expected. Results from the NGC and SGC Patchy datasets are broadly consistent, though the latter has $\sim$\,$2.5\times$ larger variance due to the smaller survey volume. Considering next the BOSS results, we find a similar story, \textit{i.e.} a noisy dataset with generally increasing amplitudes on small scales. Overall, the observational dataset appears consistent with the distribution of Patchy mocks; unlike for the 2PCF and 3PCF \citep{2016MNRAS.456.4156K}, the Patchy mocks are not calibrated using the 4PCF data, thus the agreement in the 4PCF is not necessarily obvious \textit{a priori}. Due to the high correlations between individual 4PCF measurements, it is difficult to assert a strong detection of non-Gaussianity from a visual inspection of the BOSS dataset; rather, we look to a statistical analysis, as described below. 

\begin{figure}[!ht]
    \subfloat[Variance Comparison\label{subfig: variance-comparison}]{%
      \includegraphics[width=0.6\textwidth]{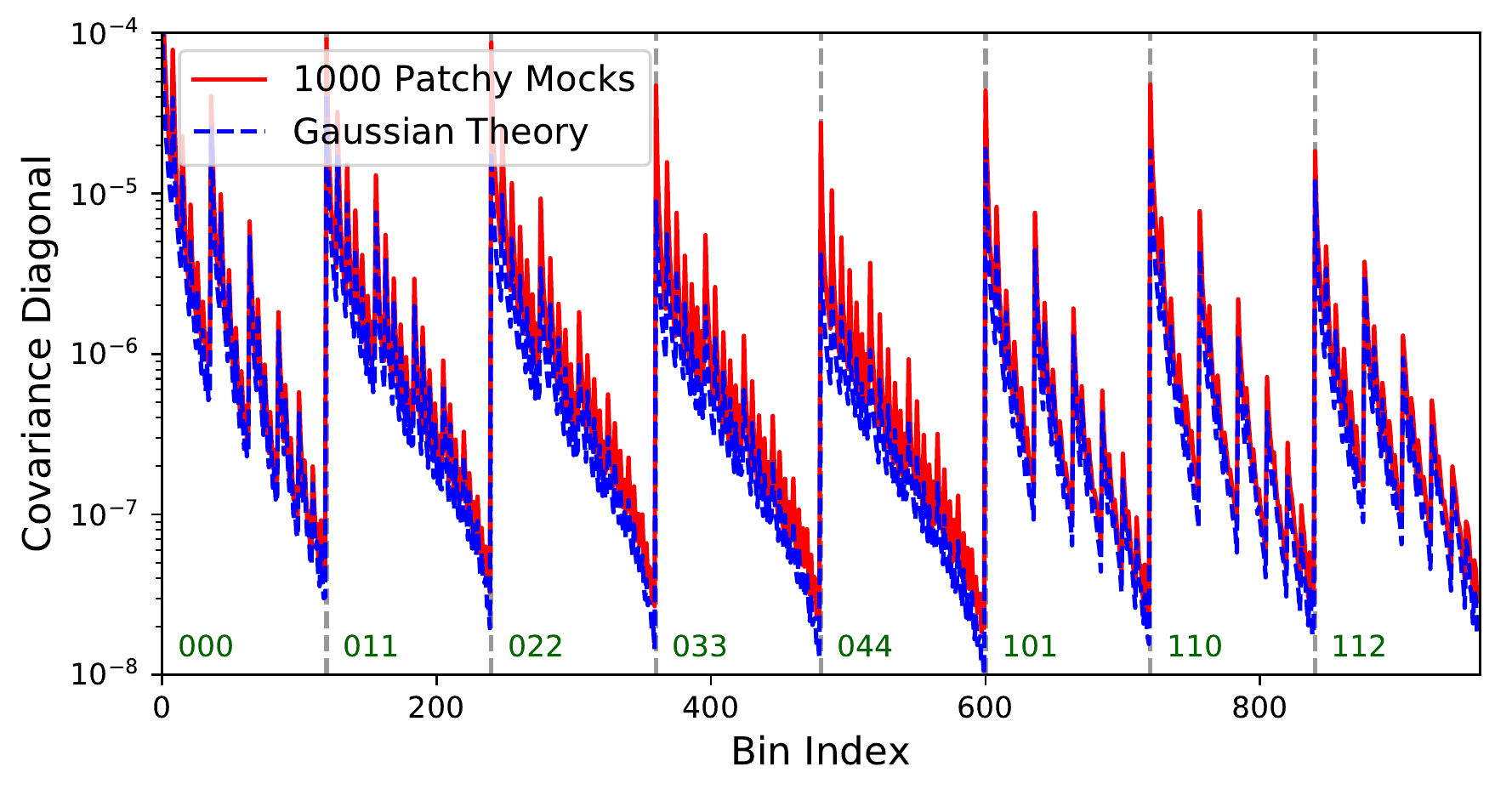}
    }
    \hfill
    \subfloat[Correlation Comparison\label{subfig: correlation-comparison}]{%
      \includegraphics[width=\textwidth]{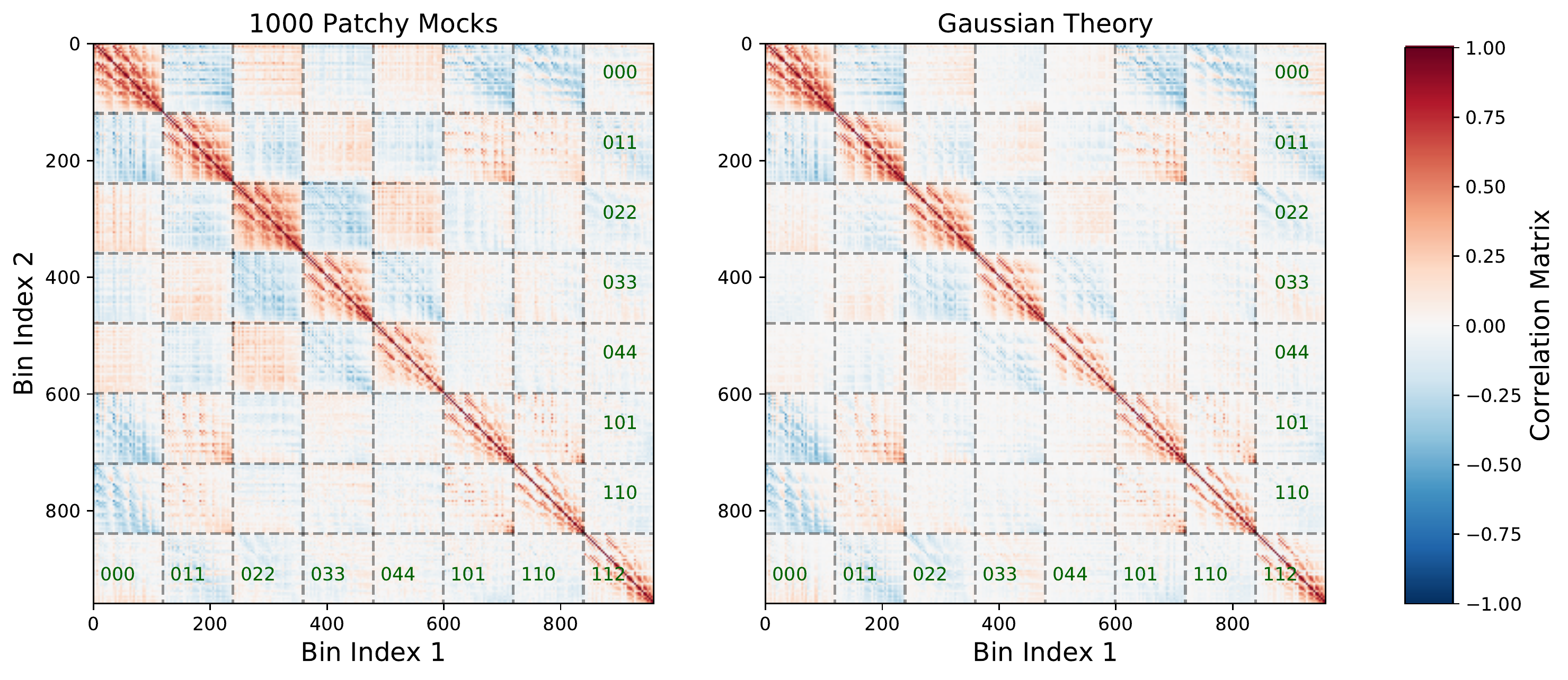}
    }
    \caption{Comparison of sample and theoretical covariance matrices for the connected 4PCF used in this work. We show results only for the NGC region, with sample covariances computed from \eqref{eq: sample-cov} using 1\,000 Patchy mocks. The theoretical covariances are those of \citep{npcf_cov} and assume isotropy, Gaussianity and a periodic-box geometry. Fig.\,\ref{subfig: variance-comparison} shows the diagonal elements of the covariance, whilst Fig.\,\ref{subfig: correlation-comparison} compares the correlation matrices, defined as the covariance normalized by its diagonal (\textit{i.e.} $\mathsf{C}_{ij}/\sqrt{\mathsf{C}_{ii} \mathsf{C}_{jj}}$). Each submatrix refers to a different set of multiplets $\{\ell_1,\ell_2,\ell_3\}$, labelled in green, and within the submatrices, the elements are ordered in increasing radius as in Fig.\,\ref{fig: boss-4pcfs}. We show only the first 8 out of 42 multiplets for clarity. We note generally good agreement between the matrices, though some differences are evident on the diagonal. Since the theoretical covariance is only used as a tool to facilitate efficient data compression, such differences do not bias the analysis of this work.}
    \label{fig: cov-comparison}
\end{figure}

In Fig.\,\ref{fig: cov-comparison} we show the variance and correlation matrix of the connected 4PCF, measured from the 1\,000 Patchy simulations using \eqref{eq: sample-cov}. As for the lower-point functions, the variance is a strongly declining function of scale (comparing with the bin-centers shown in Fig.\,\ref{fig: boss-4pcfs}); however, if one multiplies by $r_1r_2r_3$ (as in Fig.\,\ref{fig: boss-4pcfs}), it is roughly scale-independent. This reaffirms our conclusion that non-Gaussianity can be best probed on small scales (\textit{i.e.} using tetrahedrons with at least one short side), where the signal is largest. From the correlation matrices (equal to the covariances normalized by their diagonal), we see strong off-diagonal contributions approaching unity for bins within the same $\{\ell_1,\ell_2,\ell_3\}$ multiplet. This is again visible in Fig.\,\ref{fig: boss-4pcfs} in the coherent fluctuations between neighboring bins, and is a consequence of the high dimensionality of the 4PCF statistic. Significant (anti-)correlations are also seen between neighboring multiplets (\textit{i.e.} those with $\ell$'s differing by at most one), which is a generic feature of our basis, and is similar to the structure found for the 3PCF in \citep{2017MNRAS.468.1070S}. 

Additionally shown in Fig.\,\ref{fig: cov-comparison} is a comparison between the Patchy sample covariance and the theoretical covariance discussed in \S\ref{subsec: covariance}. From the diagonal elements, we see that the variance is fairly well reproduced by the Gaussian model, though the ratio of sample to theory covariance is roughly $ 1.7\pm0.6$, showing an overall underprediction from the theory. More troublingly, this ratio is observed to vary with $\ell$; for example, results from the $\{1,1,0\}$ multiplet match the sample covariance well, but those from the $\{0,4,4\}$ multiplet are a clear underprediction. On the other hand, the correlation plots indicate that the matrix structure is well reproduced, including the non-trivial correlations within multiplets and between them. In \citep{npcf_cov}, it was shown that the 4PCF covariance \textit{was} a good model for simulations with periodic-box geometries; it is therefore likely that the discrepancies arise due to the impact of the non-uniform survey geometry. Such effects are known to be of importance in the covariances of lower-order statistics \citep[e.g.,][]{2020PhRvD.102l3517W,2020PhRvD.102l3521W,2016MNRAS.462.2681O,2020MNRAS.491.3290P,2019MNRAS.490.5931P}, thus this is unsurprising. If the theoretical covariances were being used directly in the analysis this would be a cause for concern (as shown in Appendix \ref{appen: rescaled-theory-cov}), but is not in our context, since the theory covariance is used only to define the compression vectors (\S\ref{subsec: compression}).

\begin{figure}
    \centering
    \includegraphics[width=\textwidth]{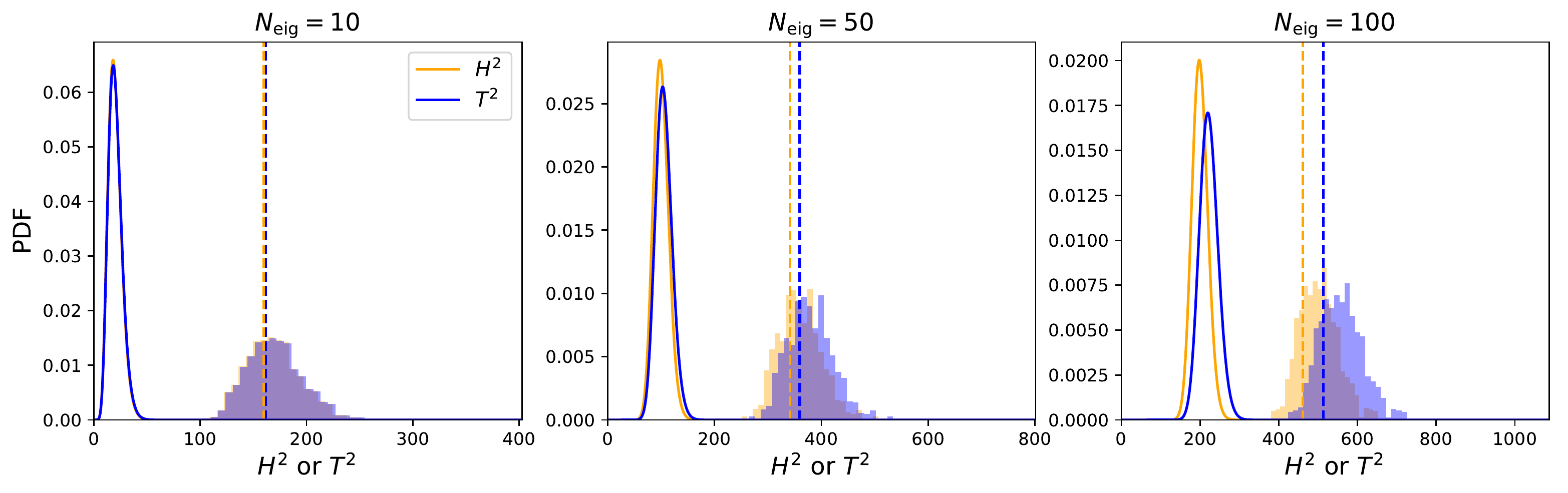}
    \caption{PDFs of the test statistics discussed in \S\ref{sec: techniques}, showing the significance of a detection of non-zero 4PCF in the BOSS CMASS dataset as a function of the number of basis vectors in the compression scheme, $N_{\rm eig}$. In orange and blue we show the $H^2$ and $T^2$ statistics respectively, both of which are related to the $\chi^2$ statistic, but differently account for the finite number of mocks. The $H^2$ statistic follows a $\chi^2$-distribution, which will show a reduced width as $N_{\rm eig}$ approaches $N_{\rm mocks}$. Sample covariances are computed by jackknifing the 1\,000 Patchy mocks (using each subset of 999 mocks to define a sample covariance), and results are shown for four choices of compressed data-vector size, with the BOSS data shown as vertical dashed lines. The sample statistics of the NGC and SGC regions are summed, giving a resulting distribution that can be evaluated via a convolution (as discussed in the main text). In each case, we report a strong detection of non-Gaussianity, at the level equivalent to a Gaussian significance of $8.1\sigma, 8.2\sigma$ and $8.1\sigma$ for $N_{\rm eig} = 10$, $50$, and $100$ respectively, using the $T^2$ statistic.}
    \label{fig: boss-4pcf-detection}
\end{figure}

Given the connected 4PCF measurements and covariances, we may proceed to assess the significance of a detection of non-Gaussianity. As in \S\ref{subsec: compression}, we first compress the data by projecting it onto a basis of eigenmodes of the theory covariance. Here, we perform the compression for the NGC and SGC regions separately (assuming them to be uncorrelated), using $N_{\rm eig} = 10, 50,$ or $100$. The test statistics $H^2$ (the Hartlap-rescaled $\chi^2$ of \ref{eq: hartlap-chi2}) and $T^2$ (from \ref{eq: compressed-chi2}) are computed both for the BOSS data and the Patchy mocks (obtaining the relevant covariance matrices via jackknifing in the latter case), then compared to the theory distributions. To combine the two data chunks, we simply add the $H^2$ or $T^2$ sample statistics, such that their expected distributions (under the null hypothesis) become the convolution of two $\chi^2$ distributions or of two copies of \eqref{eq: T2-pdf} (both with $N_{\rm eig}$ degrees of freedom). These are easily evaluated using Fast Fourier Transforms of the single-sample PDFs. For small $N_{\rm eig}$, we expect both likelihoods to yield similar results, though the less-conventional $T^2$ statistic is the more appropriate choice as $N_{\rm eig}$ approaches $N_{\rm mocks}$ \citep{2016MNRAS.456L.132S}. 

Fig.\,\ref{fig: boss-4pcf-detection} shows the test statistics obtained from the Patchy and BOSS data compared against the distribution expected under the null hypothesis of zero connected 4PCF. The conclusion is readily apparent; \textbf{we report a strong detection of non-Gaussianity in the 4PCF of BOSS}. By comparing the BOSS $T^2$ statistic to the CDF of the null distribution, we find a probability-to-exceed of $\sim$\,$10^{-15}$; \textbf{this corresponds to a Gaussian detection significance of $\sim$\,$8\sigma$}. As $N_{\rm eig}$ increases, the detection significances remain consistent; this indicates that our compression scheme has captured the majority of the signal-to-noise, and that we are not yet limited by the number of mocks. Extending to $N_{\rm eig} = 250$, the detection significance reduces to $\sim$\,$7\sigma$, due to the limited number of mocks. We further note that the BOSS and Patchy results are highly consistent, as observed in Fig.\,\ref{fig: boss-4pcfs}. The results are broadly consistent using the Hartlap-rescaled statistic $H^2$ and its $\chi^2$-distribution, though the detection significance increases to $\sim$\,$8.5\sigma$ with $N_{\rm eig}=250$, most likely due to the artificially narrow theory distribution. We caution however that non-Gaussianity of the 4PCF likelihood will modify the exact detection significances somewhat (cf.\,\S\ref{subsec: non-gaussian-likelihood}).

\begin{figure}
    \centering
    \includegraphics[width=\textwidth]{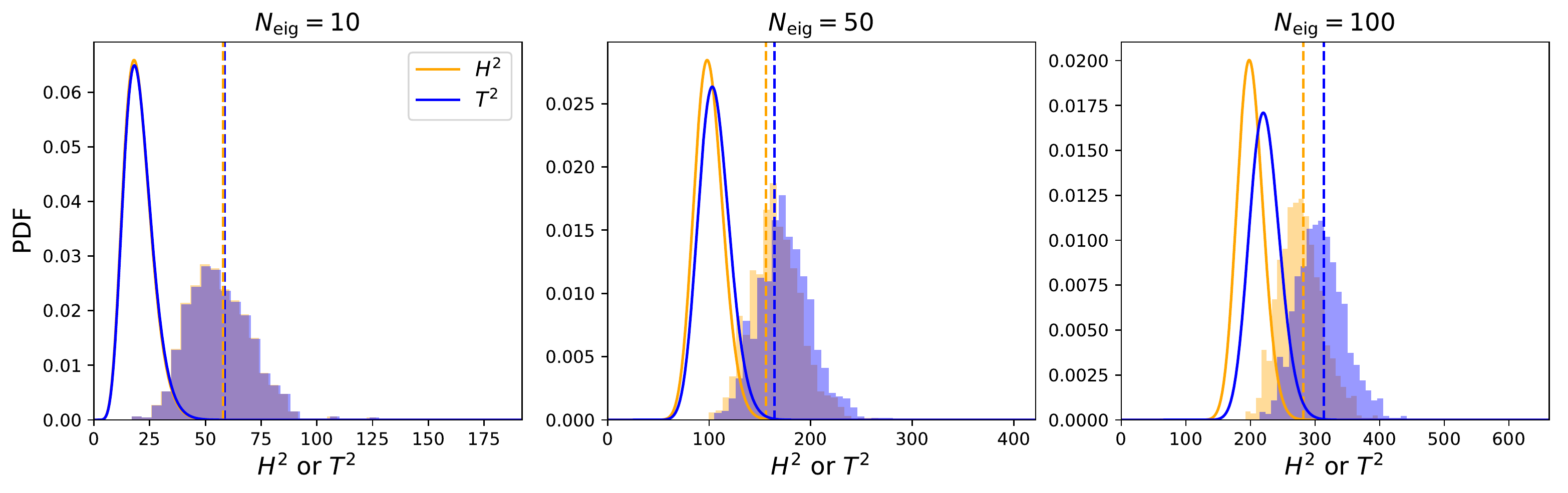}
    \caption{As Fig.\,\ref{fig: boss-4pcf-detection}, but enforcing a minimum separation of $\Delta r = 14\,\Mpch$ between any two galaxies forming the 4PCF tetrahedra. In this case, we find the detection strength is equivalent to a Gaussian significance of $4.7\sigma$, $4.0\sigma$ and $4.0\sigma$ for $N_{\rm eig} = 10, 50$, and $100$ respectively. As expected, this significance is less than that of Fig.\,\ref{fig: boss-4pcf-detection}, since the non-Gaussian signature is found primarily on short-scales. However, this result provides a practical sense of the \textit{useful} 4PCF signal strength, since the modes expected to be most difficult to model have been removed.}
    \label{fig: cut-boss-4pcf-detection}
\end{figure}

As observed in Fig.\,\ref{fig: boss-4pcfs}, most of the signal-to-noise in the connected 4PCF occurs on small scales, where non-Gaussianity is largest. Whilst our analysis has restricted to $r_{\rm min} = 20\,\Mpch$, this does not in fact enforce a minimum separation on \textit{all} points in the 4PCF tetrahedron, just the distance of the secondary galaxies from some primary vertex. As an example, a 4PCF bin with $r_1$, $r_2$ and $r_3$ legs in the ranges $[20,34)\,\Mpch$, $[34,48)\,\Mpch$ and $[48,62)\,\Mpch$ (corresponding to the first bin in Fig.\,\ref{fig: boss-4pcfs}), can have arbitrarily small (though non-zero) values of $|\vr_1-\vr_2|$ and $|\vr_2-\vr_3|$. Whilst not a problem for the detection of the 4PCF, such modes are challenging to model since any perturbative description necessarily breaks down beyond the non-linear scale and there will be significant effects from baryonic physics not included in simulations. To this end, we consider restricting to bins with centers separated by greater than $\Delta r = 14\,\Mpch$; this enforces a minimum separation of $\Delta r$ between each galaxy in the 4PCF tetrahedron, thus avoiding the most difficult-to-model modes. It additionally reduces the dimensionality by a factor of roughly two. The corresponding detection PDFs are shown in Fig.\,\ref{fig: cut-boss-4pcf-detection}, and are heuristically similar to those of Fig.\,\ref{fig: boss-4pcf-detection}. Here, we find a probability-to-exceed of $\sim$\,$10^{-5}$, equivalent to a Gaussian significance of $\sim$\,$4.7\sigma$ for $N_{\rm eig} = 10$. Whilst this significance is noticeably decreased from that of Fig.\,\ref{fig: boss-4pcfs}, it is still a clear detection, implying that \textbf{a non-zero 4PCF can still be detected in the BOSS dataset when restricting to scales outside the non-linear regime.}

\section{Summary and Conclusions}\label{sec: summary}

To maximize the information gain from future spectroscopic surveys, it is vital to harness statistics that extract maximal information from the galaxy density field. Whilst analyses have historically focused around the 2PCF or power spectrum, there has been a recent effort to extract information from the higher-point functions, in particular the 3PCF or bispectrum \citep{2018MNRAS.478.4500P,2020JCAP...05..005D,2017MNRAS.465.1757G,2017MNRAS.469.1738S}. In this work, we present a practical method for measuring the next-order statistic, the 4PCF, and quantify its signal-to-noise in the BOSS dataset. In particular, we extend the NPCF algorithms of \citep{npcf_algo,npcf_generalized} to measure the \textit{connected} 4PCF by removing the Gaussian piece (which is degenerate with the 2PCF) at the estimator level. This is unlike previous works, and ensures that our measurement is specifically one of non-Gaussianity, rather than a recapitulation of known physics. The estimator is fast (scaling quadratically with the galaxy number density), corrected for the non-uniform survey geometry, and implemented in the public \textsc{encore} code. We verify its performance on a suite of lognormal simulations at high redshift, before applying it to the BOSS dataset and Patchy simulations. Analysis of the higher-point functions is hampered by their high dimension; here, we implement a signal-to-noise-based compression scheme (originally proposed in \citep{2000ApJ...544..597S}), which allows us to project the 4PCF into a set of $\sim 50$ numbers with minimal impact on the detection significance. Performing a classical $\chi^2$-like analysis in the compressed subspace leads to an $8.1\sigma$ detection of the non-Gaussian 4PCF, or $4.7\sigma$ if we restrict to galaxy quadruplets with separations outside the highly non-linear regime.

Based on this work, a number of avenues present themselves for future study:
\begin{itemize}
    \item \textbf{Theory Model}: To extract information from the 4PCF, we require a physical model for its dependence on cosmological and galaxy formation parameters. Whilst this is not trivial to compute, it is tractable, at least at tree level. Simulation-based approaches, such as emulators \citep[e.g.,][]{2020PhRvD.102f3504K}, also provide an interesting prospect for obtaining theory predictions in difficult-to-model regimes. 
    \item \textbf{Fisher Forecasting}: In the absence of a theory model, one can test the information content of the 4PCF with simulation-based Fisher forecasts. This requires a large number of simulations of varying cosmological and hydrodynamic parameters \citep[e.g.,][]{2021JCAP...04..029H}, which may be analyzed using a fast algorithm such as \textsc{encore}.
    \item \textbf{Odd-Parity 4PCF}: The present work has considered only 4PCF multiplets with even $\ell_1+\ell_2+\ell_3$, \textit{i.e.} the contribution invariant under parity transformations. Complementary to this is the set of multiplets with odd $\ell_1+\ell_2+\ell_3$, which can be measured in a similar fashion, and probes parity-violating physics. This will be discussed in \citep{4pcf_odd}. 
    \item \textbf{Anisotropic 4PCF}: Due to redshift-space distortions, the density field is statistically anisotropic, with azimuthal symmetry about the line of sight. As such, the full 4PCF contains both an isotropic contribution (discussed in this work) and an anisotropic component. The latter can be probed in a similar manner to this work (using the estimators derived in \citep{npcf_generalized}), and will allow all the 4PCF information to be captured.
    \item \textbf{4PCF Likelihood}: A major assumption in this work is that the 4PCF is drawn from some multivariate Gaussian distribution. As shown in a number of works \citep[e.g.,][]{2012PhRvD..86f3009S,2019MNRAS.485.2956H,2019PhRvD.100l3511M}, this assumption is questionable and should be examined further, particularly if one wishes to use the 4PCF to derive constraints on cosmological parameters.
    \item \textbf{Observational Systematics}: Since the present work has focused on the detectability of the non-Gaussian 4PCF signal, we have assumed that the standard galaxy weighting scheme (\S\ref{sec: data}) can sufficiently account for systematic effects in the data. To derive parameter constraints from our signal, a more careful treatment of such effects, in particular those of fiber collisions, is likely needed. 
\end{itemize}

Given the significance with which we can detect the 4PCF in current datasets, it is worth considering prospects for future surveys. Whilst the cosmic volume mapped will grow dramatically with efforts such as DESI and Euclid, the majority of the galaxy observations will take place at higher redshifts, where the relative importance of the 4PCF will be severely diminished.\footnote{Whilst there will be a number of low-redshift surveys (e.g., the DESI Bright Galaxy Survey (BGS); \citep{2016arXiv161100036D}), these will be mostly limited to small volumes, reducing their constraining power.} In particular, the squared signal-to-noise of a 4PCF detection scales with redshift approximately as $D^4(z)$ for linear growth rate $D(z)$,\footnote{This occurs since the tree-level trispectrum scales as $P_L^3(k,z)\propto D^6(z)$ and the covariance as $P_L^4(k,z)\propto D^8(z)$, ignoring bias evolution} and linearly with volume. Using the effective volume and redshift parameters from \citep[Table\,2.3]{2016arXiv161100036D} and \citep[Table\,1]{2019JCAP...11..034C}, a back-of-the-envelope calculation suggests that the overall 4PCF squared signal-to-noise will increase by a factor of $\sim$\,$60\%$ and $\sim$\,$100\%$ for DESI and Euclid respectively, assuming the tracers to be similar. This is significantly below that which one might expect from pure volume-rescaling arguments, and is caused by the strong redshift dependence. That said, the detection is still likely to be strong, and consequently one may expect the 4PCF to become a useful probe of physical parameters, both those pertaining to $\L$CDM and to more exotic phenomena such as primordial non-Gaussianity and modified gravity.

\begin{acknowledgments}
\footnotesize
It is a pleasure to thank Robert Cahn, Daniel Eisenstein, Simone Ferraro, Alex Krolewski, Moritz M\"unchmeyer, Ue-Li Pen, David Spergel, and Matias Zaldarriaga for stimulating discussions. We are additionally grateful to Robert Cahn, Regina Demina, and David Spergel for insightful comments on a draft of this manuscript. OHEP acknowledges funding from the WFIRST program through NNG26PJ30C and NNN12AA01C and thanks the Simons Foundation and the University of Florida for their support. ZS thanks Lawrence Berkeley National Laboratory for the use of affiliate resources during the period of this work.

The authors are pleased to acknowledge that the work reported on in this paper was substantially performed using the Princeton Research Computing resources at Princeton University, which is a consortium of groups led by the Princeton Institute for Computational Science and Engineering (PICSciE) and the Princeton Office of Information Technology's Research Computing Division.
\end{acknowledgments}

\appendix

\section{Gaussian Theory Model}\label{appen: gaussian-deriv}
Below, we derive the expected form of the 4PCF in the Gaussian (but non-linear) regime, \textit{i.e.} the disconnected term, as discussed in \S\ref{sec: lognormals}. Before projection, the Gaussian 4PCF is given by a set of Wick contractions of pairs of density fields:
\beq\label{eq: gaussian-4pcf-preproject}
    \zeta^\mathrm{G}(\vr_1,\vr_2,\vr_3) &=& \av{\delta(\vs)\delta(\vs+\vr_1)\delta(\vs+\vr_2)\delta(\vs+\vr_3)} = \xi(\vr_1)\xi(\vr_2-\vr_3) + \text{2 perms.}\\\nonumber
    &=& \xi(\vr_1)\int_{\vk}P(\vk)e^{i\vk\cdot(\vr_2-\vr_3)} + \text{2 perms.},
\eeq
writing the 2PCF in terms of the power spectrum in the second line. Accounting for RSD, $\xi(\vr)$ and $P(\vk)$ may be expanded as Legendre series around the (assumed fixed) line-of-sight (LoS) $\hn$:
\beq\label{eq: pk-xi-decomp}
    \xi(\vr) = \sum_{L}\xi_L(r)L_L(\hr\cdot\hn), \quad P(\vk) = \sum_L P_L(k)L_L(\hk\cdot\hn),
\eeq
where $L_L(\mu)$ is a Legendre polynomial of order $L$, which is even if the overdensity field conserves parity. In the simplest instance of Kaiser linear theory \citep{1987MNRAS.227....1K}, only moments with $L\in\{0,2,4\}$ are non-zero. Inserting \eqref{eq: pk-xi-decomp} into \eqref{eq: gaussian-4pcf-preproject} gives
\beq\label{eq: 4pcf-gaussian-pre-proj}
    \zeta^{\rm G}(\vr_1,\vr_2,\vr_3) &=& \sum_{LL'}\xi_L(r_1)\int_{\vk}P_{L'}(k)e^{i\vk\cdot(\vr_2-\vr_3)}\int \frac{d\hn}{4\pi}L_L(\hr_1\cdot\hn)L_{L'}(\hk\cdot\hn) + \text{2 perms.}\\\nonumber 
    &=& \sum_L\frac{1}{2L+1}\,\xi_L(r_1)\int_{\vk}P_L(k)e^{i\vk\cdot(\vr_2-\vr_3)}L_L(\hr_1\cdot\hk),
\eeq 
where we have additionally inserted an angular integral over the LoS, justified since we consider only the isotropic 4PCF, which cannot depend on $\hn$.\footnote{More formally, since the basis functions are isotropic, they must satisfy $\P_{\ell_1\ell_2\ell_3}(\mathcal{R}\hr_1,\mathcal{R}\hr_2,\mathcal{R}\hr_3) = \P_{\ell_1\ell_2\ell_3}(\hr_1,\hr_2,\hr_3)$ for any rotation $\mathcal{R}$. After projection onto the isotropic basis, we can perform an arbitrary rotation of the 4PCF, which is equivalent to integrating over the possible LoS directions.} In the final line, we perform the $\hn$ integral via Legendre polynomial orthogonality \citep[\S14.7.6]{nist_dlmf}. Expanding the exponentials $e^{i\vk\cdot\vr_2}$ and $e^{-i\vk\cdot\vr_3}$ via the Rayleigh plane wave identity \citep[Eq.,\,16.63]{2005mmp..book.....A} gives
\beq
    \int \frac{d\hn}{4\pi}\,\xi(\vr_1)\xi(\vr_2-\vr_3) &=& \sum_{L,\ell_1,\ell_2}\frac{(4\pi)^3}{(2L+1)^2}\,i^{\ell_1-\ell_2}\xi_L(r_1)\int_{\vk}P_L(k)j_{\ell_1}(kr_2)j_{\ell_2}(kr_3)\\\nonumber
    &&\,\times\,Y_L^M(\hk)Y_{\ell_1}^{m_1}(\hk)Y_{\ell_2}^{m_2}(\hk)Y_L^{M*}(\hr_1)Y_{\ell_1}^{m_1*}(\hr_2)Y_{\ell_2}^{m_2*}(\hr_3),
\eeq
additionally rewriting the Legendre polynomial in terms of spherical harmonics \citep[\S14.30.9]{nist_dlmf} and introducing spherical Bessel functions $j_L(x)$. The angular integral over $\hk$ yields a Gaunt integral \citep[\S34.3.22]{nist_dlmf}; writing this explicitly in terms of 3-$j$ symbols gives
\beq
    \int \frac{d\hn}{4\pi}\,\xi(\vr_1)\xi(\vr_2-\vr_3) &=& \sum_{L\ell_1\ell_2}\left[\frac{(4\pi)^3(2\ell_1+1)(2\ell_2+1)}{(2L+1)^3}\right]^{1/2}\tjo{L}{\ell_1}{\ell_2}i^{\ell_1-\ell_2}\xi_L(r_1)\\\nonumber
    &&\,\times\,\int\frac{k^2dk}{2\pi^2}P_L(k)j_{\ell_1}(kr_2)j_{\ell_2}(kr_3)\sum_{Mm_1m_2}\tj{L}{\ell_1}{\ell_2}{M}{m_1}{m_2}Y_L^{M*}(\hr_1)Y_{\ell_1}^{m_1*}(\hr_2)Y_{\ell_2}^{m_2*}(\hr_3).
\eeq
We recognize the sum over three $m$ indices as the basis function $\P_{L\ell_1\ell_2}(\hr_1,\hr_2,\hr_3)$ in \eqref{eq: N=4-basis} (assuming even $L+\ell_1+\ell_2$, such that $\P_{\ell_1\ell_2\ell_3}^* = \P_{\ell_1\ell_2\ell_3}$). By orthogonality, we can read off the multipole coefficients as
\beq\label{eq: gaussian-4pcf-tmp}
    \zeta^\mathrm{G}_{L\ell_1\ell_2}(r_1,r_2,r_3) &=& i^{\ell_1-\ell_2}\left[\frac{(4\pi)^3(2\ell_1+1)(2\ell_2+1)}{(2L+1)^3}\right]^{1/2}\tjo{L}{\ell_1}{\ell_2}\xi_L(r_1)\int\frac{k^2dk}{2\pi^2}P_L(k)j_{\ell_1}(kr_2)j_{\ell_2}(kr_3),
\eeq
with analogous results found for the other two permutations. This has support for all multiplets in which the triangle conditions on $\{L,\ell_1,\ell_2\}$ are satisfied.

Defining 
\beq
    f^{L}_{\ell\ell'}(r,r') = \int\frac{k^2dk}{2\pi^2}P_L(k)j_{\ell}(kr)j_{\ell'}(kr'),
\eeq
with the special case $f^L_{\ell0}(r,0) = i^{-\ell}\xi_\ell(r)$, \eqref{eq: gaussian-4pcf-tmp} can be written succinctly as
\beq\label{eq: gaussian-4pcf-full-appen}
    \zeta^\mathrm{G}_{\ell_1\ell_2\ell_3}(r_1,r_2,r_3) &=& \left[\frac{(4\pi)^3(2\ell_2+1)(2\ell_3+1)}{(2\ell_1+1)^3}\right]\tjo{\ell_1}{\ell_2}{\ell_3}i^{\ell_2-\ell_3}\xi_{\ell_1}(r_1)f^{\ell_1}_{\ell_2\ell_3}(r_2,r_3) + \text{2 perms.}
\eeq

An interesting case is the limit of zero RSD, in which case $\xi_\ell$ and $P_\ell$ are zero for $\ell>0$. Following simplification, this yields only three non-zero multiplets:
\beq\label{eq: iso-gaussian-4pcf}
    \zeta_{0\ell\ell}^\mathrm{G,no\,RSD}(r_1,r_2,r_3) &=& (4\pi)^{3/2}(-1)^{\ell}\sqrt{2\ell+1}\xi_0(r_1)f^0_{\ell\ell}(r_2,r_3),\\\nonumber
    \zeta_{\ell0\ell}^\mathrm{G,no\,RSD}(r_1,r_2,r_3) &=& (4\pi)^{3/2}(-1)^{\ell}\sqrt{2\ell+1}\xi_0(r_2)f^0_{\ell\ell}(r_1,r_3),\\\nonumber
    \zeta_{\ell\ell0}^\mathrm{G,no\,RSD}(r_1,r_2,r_3) &=& (4\pi)^{3/2}(-1)^{\ell}\sqrt{2\ell+1}\xi_0(r_3)f^0_{\ell\ell}(r_1,r_2).
\eeq
A detection of non-zero 4PCF in any multiplet not of this form is thus evidence for RSD effects (in the disconnected piece) or non-Gaussianity (in the connected piece).

Finally, we consider radial binning. As in \S\ref{sec: background}, the 4PCF algorithm natively measures the correlators averaged across bins of finite width, thus the same averaging must be performed for the theory model. Given a bin triplet $\{a,b,c\}$, the binned theory is defined by
\beq
    \zeta^{abc}_{\ell_1\ell_2\ell_3} = \frac{(4\pi)^3}{v_av_bv_c}\int r_1^2dr_1\,\Theta^a(r_1)\int r_2^2dr_2\,\Theta^b(r_2)\int r_3^2dr_3\,\Theta^c(r_3)\,\zeta_{\ell_1\ell_2\ell_3}(r_1,r_2,r_3),
\eeq
where $v_u = 4\pi\int r^2dr\,\Theta^u(r)$. In the above model, binning can be simply realized by replacing $\xi_\ell$ and $f_{\ell_1\ell_2}^L$ by their binned equivalents, \textit{i.e.}
\beq
    \zeta^{\mathrm{G},abc}_{\ell_1\ell_2\ell_3} &=& (4\pi)^{3/2}\sqrt{(2\ell_1+1)(2\ell_2+1)(2\ell_3+1)}\tjo{\ell_1}{\ell_2}{\ell_3}\\\nonumber
    &&\,\times\,\left\{\frac{i^{\ell_2-\ell_3}}{(2\ell_1+1)^2}\,\xi^a_{\ell_1}f^{\ell_1,bc}_{\ell_2\ell_3}+\frac{i^{\ell_1-\ell_3}}{(2\ell_2+1)^2}\,\xi^b_{\ell_2}f^{\ell_2,ac}_{\ell_1\ell_3}+\frac{i^{\ell_1-\ell_2}}{(2\ell_3+1)^2}\,\xi_{\ell_3}^c(r_3)f^{\ell_3,ab}_{\ell_1\ell_2}\right\},
\eeq
where
\beq\label{eq: xi-ell,f-ell-binned}
    \xi_\ell^u = i^\ell \int \frac{k^2dk}{2\pi^2}P_\ell(k)j_\ell^u(k), \qquad f_{\ell\ell'}^{L,uv} = \int\frac{k^2dk}{2\pi^2}P_L(k)j_\ell^u(k)j_{\ell'}^v(k).
\eeq
These use the bin-averaged spherical Bessel functions, defined by
\beq\label{eq: bin-averaged-sBF}
    j_\ell^u(k) = \frac{4\pi}{v_u}\int r^2dr\,j_\ell(kr)\Theta^u(r).
\eeq
These are in fact analytic, and their general form can be expressed in terms of generalized hypergeometric functions as in \citep[Appendix A][]{2020MNRAS.492.1214P}. For integer $\ell$, they are simple sums of trigonometric functions and the Sine integral, and easily evaluated using e.g.,, \textsc{Mathematica}. Performing such computations analytically allows for the theory model to be evaluated quickly and at high precision.

\section{Analysis Using a Fitted Theory Covariance Matrix}\label{appen: rescaled-theory-cov}

\begin{figure}
    \centering
    \includegraphics[width=0.45\textwidth]{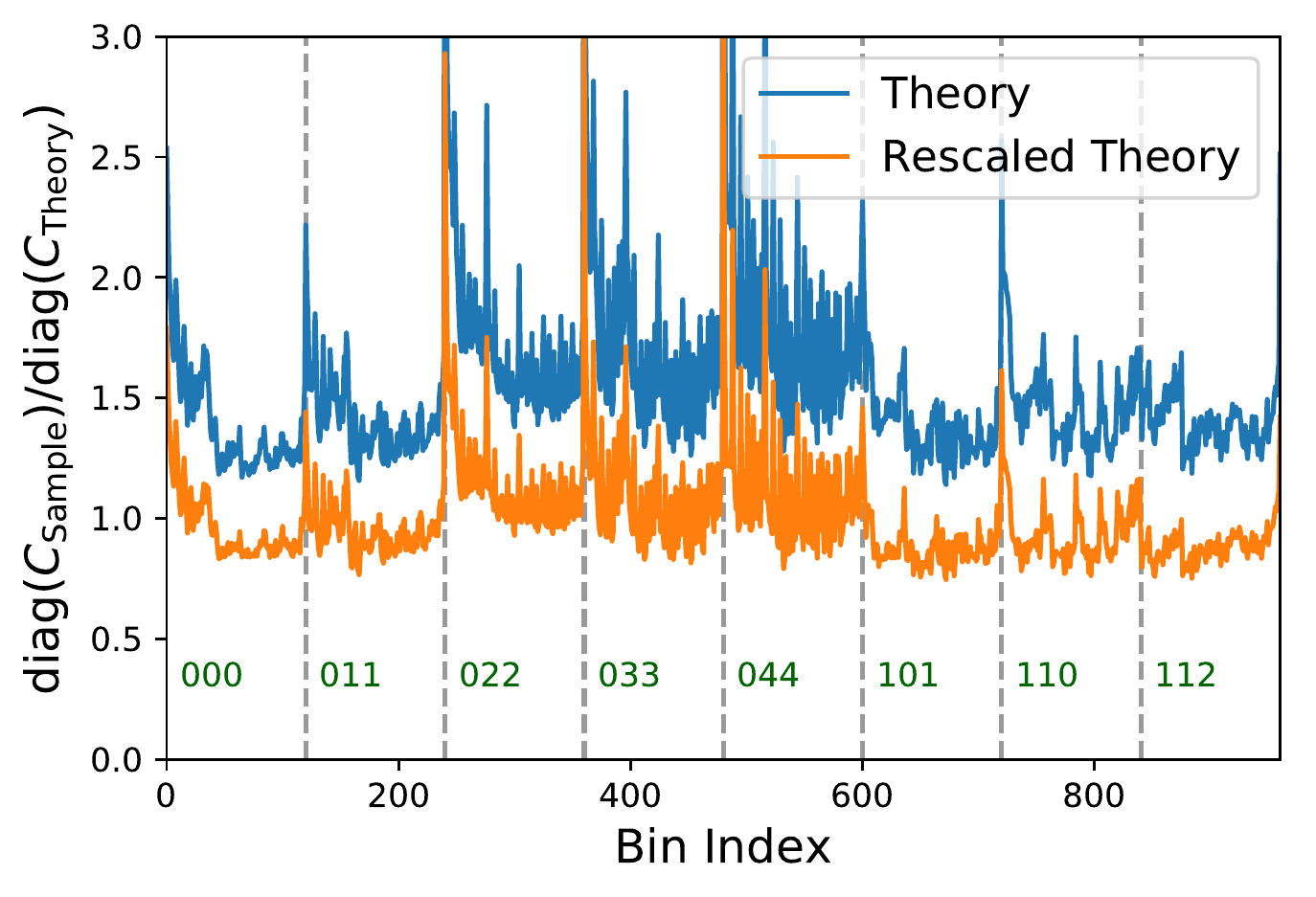}
    \caption{Comparison between the variances of the Patchy mocks and those obtained from the theoretical covariance model discussed in \S\ref{subsec: covariance}. We present results both before and after fitting for the volume $V_{\rm eff}$ and shot-noise $\bar{n}_{\rm eff}$ by comparing the theoretical and sample covariances. Results are shown only for the first 8 of 42 multiplets, as in Fig.\,\ref{fig: cov-comparison}, again restricting to the NGC data chunk. The inclusion of fitting parameters leads to better agreement between theory and simulations, though not perfect, as evidenced by the underpredictions at small radii.}
    \label{fig: rescaled-variance}
\end{figure}

As demonstrated in Fig.\,\ref{fig: cov-comparison}, the Gaussian covariance derived in \citep{npcf_cov} is not in perfect agreement with the sample covariance from the Patchy mocks, particularly when the variance is considered. As advocated for in \citep{npcf_cov}, one way to ameliorate this is to adjust the effective survey volume $V$ and shot-noise parameter $1/\bar{n}$, which can, to some extent, mimic the effects of non-uniform geometry and non-Gaussianity \citep[cf.,][]{2016MNRAS.462.2681O,2019MNRAS.487.2701O,2020MNRAS.491.3290P}. Here, we implement this approach, fitting for $V$ and $1/\bar{n}$ by comparing the theoretical and sample covariances using the Kullback-Leibler divergence as in \citep{2016MNRAS.462.2681O}, which accounts for the expected (Wishart) distribution of the mock-based covariance. For the NGC data chunk, minimization leads to the best-fit parameters $\{V_{\rm eff} = 1.57h^{-3}\mathrm{Gpc}^3, \bar{n}^{-1}_{\rm eff} = 4.2\times 10^3\,h^3\mathrm{Mpc}^{-3}\}$, which differ significantly from the fiducial values $\{V = 1.9h^{-3}\mathrm{Gpc}^3, \bar{n}^{-1} = 3.0\times 10^3\,h^3\mathrm{Mpc}^{-3}\}$. Comparison of the sample and theoretical variances, both before and after parameter fitting, are shown in Fig.\,\ref{fig: rescaled-variance}. Noticeably, we find much improved agreement between Patchy and the theory model after the rescaling, implying that the free parameters have been able to account for some of the survey geometry and non-Gaussian effects. That said, there remain some systematic offsets in the rescaled theory covariance, particularly in bins corresponding to small radii.

\begin{figure}
    \centering
    \includegraphics[width=0.95\textwidth]{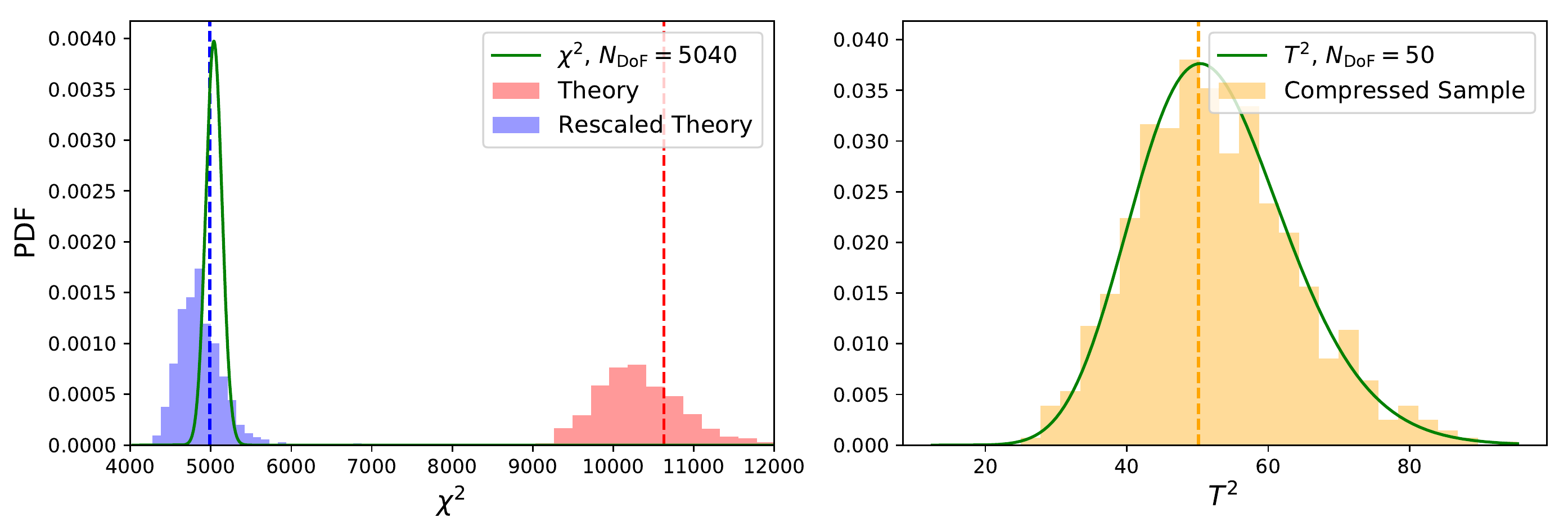}
    \caption{Distributions of the $\chi^2$ (left) and compressed $T^2$ (right) statistics from a set of mock 4PCFs with the mean subtracted. As before, the vertical lines give the corresponding results for BOSS. In the left panel, we show the $\chi^2$ statistic, computed using the theory covariance of \S\ref{subsec: compression} (red), optionally with fitted $V$ and $\bar{n}$ parameters (blue, cf.\,Fig.\,\ref{fig: rescaled-variance}). Since the signal has been removed from the data, we should see agreement between the expected distribution (green) and the empirical histograms (blue and red) if the theoretical covariances are accurate; this is clearly not the case, both before and after fitting. In the right panel, we perform an analysis in the compressed subspace using $N_{\rm eig} = 50$ basis vectors (cf.\,\S\ref{subsec: compression}), plotting the $T^2$ statistic of \eqref{eq: compressed-chi2} and the expected distribution of \eqref{eq: T2-pdf}. In this case, we do not find any evidence for bias, as expected.}
    \label{fig: no-signal-test}
\end{figure}

To test the rescaled covariance matrix more rigorously, we consider a scenario in which the expected 4PCF signal is zero, easily obtained by taking the connected 4PCF results from Patchy and subtracting the mean. This is of use since the distribution is known; if the theoretical covariances match the true underlying covariance (and the underlying likelihood is Gaussian), then the $\chi^2$ statistic
\beq
    \left.\chi^2\right|_{\rm theory} = \tilde\zeta^T\mathsf{C}^{-1}_{\rm theory}\tilde\zeta
\eeq
of the mean-subtracted data $\tilde\zeta$ will be distributed according to a $\chi^2$ distribution with $N_\zeta=5\,040$ degrees of freedom. In the left panel of Fig.\,\ref{fig: no-signal-test}, we plot the empirical $\chi^2$ distributions obtained from the $1\,000$ Patchy NGC simulations using both the fiducial and fitted theoretical covariance matrices, alongside the expected distribution. The fiducial (unrescaled) covariance leads to an empirical distribution which is highly inconsistent with that of the theory; a na\"ive analysis using this data would obtain $\chi^2/N_\zeta\approx 2$ and conclude a strong detection of non-Gaussianity, even though none was present in the data. When using the fitted covariance matrix, the result is far more reasonable (as expected, since one has effectively minimized the expectation of $|\chi^2-N_{\zeta}|$); however, the empirical and sample distributions are still not in agreement. In particular, the shape of the distribution is incorrect, and, for many data realizations, one would conclude that there was \textit{less} non-Gaussianity present than in the null-hypothesis, which is nonsensical. 

In the right panel of Fig.\,\ref{fig: no-signal-test}, we give the corresponding results for the mean-subtracted 4PCFs analyzed in the compressed subspace of \S\ref{subsec: compression} using $N_{\rm eig} = 50$ and a sample covariance computed from 1\,000 Patchy mocks. As discussed previously compression cannot lead to a spurious signal detection (assuming the 4PCF to be Gaussian distributed); this is borne out in practice, since the empirical distribution of $T^2$ is well modelled by its theory distribution \eqref{eq: T2-pdf}. The excellent agreement also indicates that the non-Gaussianity in the 4PCF \textit{likelihood} is not of significant importance here. Our main conclusion is the following: if one wishes to compute the 4PCF detection significance, it is dangerous to use the theoretical covariances directly in  $\chi^2$, since this can give significant mis-detections of the signal even when none is present. Working in the compressed subspace avoids such concerns, even though it may lead to a slight loss of signal-to-noise if the theory covariance (used to define the basis vectors) is far from the truth.

\section{Connected 4PCF Covariance}\label{appen: fully-coupled-covariance}
In \S\ref{subsec: covariance}, it was noted that the covariance matrix of the connected 4PCF contains only fully-coupled contributions \eqref{eq: basis-cov-split}. Here, we provide an explicit demonstration of this, as well as a visual comparison of the full and connected 4PCF covariance matrices using lognormal simulations.

Firstly, consider the connected estimator in the form
\beq\label{eq: connected-tmp}
    \hat\zeta^{(\rm c)}(\vr_1,\vr_2,\vr_3) &=& \frac{1}{V}\int d\vs\,\delta(\vs+\vr_0)\delta(\vs+\vr_1)\delta(\vs+\vr_2)\delta(\vs+\vr_3)\\\nonumber
    &&\,-\,\left[\frac{1}{V}\int d\vs\,\delta(\vs+\vr_0)\delta(\vs+\vr_1)\,\frac{1}{V}\int d\vs'\,\delta(\vs'+\vr_2)\delta(\vs'+\vr_3) + \text{2 perms.}\right]
\eeq
again introducing the dummy variable $\vr_0 = \vec 0$ for keeping track of the permutations. Computing the covariance of \eqref{eq: connected-tmp} and applying statistical homogeneity will give terms proportional to $V^{-\alpha}$ where $\alpha \in\{1,2,3\}$. Assuming the survey size to be much larger than the typical correlation length $r_c \approx 100\,\Mpch$, we may drop any terms with $\alpha>1$, since these are suppressed by powers of $r_c^3/V\ll 1$. As an example, consider the term arising from the expectation of two disconnected 4PCFs;
\beq
    \av{\zeta^{(\rm disc)}(\vr_1,\vr_2,\vr_3)\zeta^{(\rm disc)}(\vr_1',\vr_2',\vr_3')}&\supset&\frac{1}{V}\int d\vs_1\,\frac{1}{V}\int d\vs_1'\,\av{\delta(\vs_1+\vr_0)\delta(\vs_1'+\vr_0')}\av{\delta(\vs_1+\vr_1)\delta(\vs_1'+\vr_1')}\\\nonumber
    &&\,\times\,\frac{1}{V}\int d\vs_2\,\frac{1}{V}\int d\vs_2'\,\av{\delta(\vs_2+\vr_2)\delta(\vs_2'+\vr_2')}\av{\delta(\vs_2+\vr_3)\delta(\vs_2'+\vr_3')}\\\nonumber
    &=& \frac{1}{V}\int d\vs\,\xi(\vs+\vr_0-\vr_0')\xi(\vs+\vr_1-\vr_1')\frac{1}{V}\int d\vs'\,\xi(\vs'+\vr_2-\vr_2')\xi(\vs'+\vr_3-\vr_3'),
\eeq
relabelling variables in the second line and using Wick's theorem. This scales as $(r_c^3/V)^2$ unlike the terms in \eqref{eq: basis-cov-split}.

Decomposing the covariance into terms arising from the full and disconnected 4PCF estimators, the remaining terms in the connected covariance (\textit{i.e.} those not suppressed by additional powers of $r_c^3/V$) are given by
\beq
    \mathrm{Cov}^{(\rm full, full)} &=& \frac{1}{V}\int d\vs\,\xi(\vs+\vr_0-\vr_0')\xi(\vs+\vr_1-\vr_1')\xi(\vs+\vr_2-\vr_2')\xi(\vs+\vr_3-\vr_3')+\text{23 perms.}\\\nonumber
    &&\,+\,\xi(\vr_1-\vr_0)\xi(\vr_1'-\vr_0')\frac{1}{V}\int d\vs\,\xi(\vs+\vr_2-\vr_2')\xi(\vs+\vr_3-\vr_3') + \text{71 perms.}\\\nonumber
    \mathrm{Cov}^{(\rm full, disc)} &=& \xi(\vr_1-\vr_0)\xi(\vr_1'-\vr_0')\frac{1}{V}\int d\vs\,\xi(\vs+\vr_2-\vr_2')\xi(\vs+\vr_3-\vr_3') + \text{71 perms.}\nonumber\\
    \mathrm{Cov}^{(\rm disc, full)} &=& \xi(\vr_1-\vr_0)\xi(\vr_1'-\vr_0')\frac{1}{V}\int d\vs\,\xi(\vs+\vr_2-\vr_2')\xi(\vs+\vr_3-\vr_3') + \text{71 perms.}\nonumber\\
    \mathrm{Cov}^{(\rm disc, disc)} &=& \xi(\vr_1-\vr_0)\xi(\vr_1'-\vr_0')\frac{1}{V}\int d\vs\,\xi(\vs+\vr_2-\vr_2')\xi(\vs+\vr_3-\vr_3') + \text{71 perms.},
\eeq
dropping the $\vr_i$ labels on the LHS for brevity, and using Wick's theorem extensively. In full, we find the connected covariance 
\beq
    \mathrm{Cov}^{( c, c)} &=& \mathrm{Cov}^{(\rm full, full)} - \mathrm{Cov}^{(\rm full, disc)} - \mathrm{Cov}^{(\rm disc, full)} + \mathrm{Cov}^{(\rm disc, disc)}\\\nonumber
    &=& \frac{1}{V}\int d\vs\,\xi(\vs+\vr_0-\vr_0')\xi(\vs+\vr_1-\vr_1')\xi(\vs+\vr_2-\vr_2')\xi(\vs+\vr_3-\vr_3')+\text{23 perms.}+\mathcal{O}\left(\frac{r_c^6}{V^2}\right),
\eeq
\textit{i.e.} \eqref{eq: basis-cov-split}, including only fully-coupled terms.

\begin{figure}
    \centering
    \includegraphics[width=0.48\textwidth]{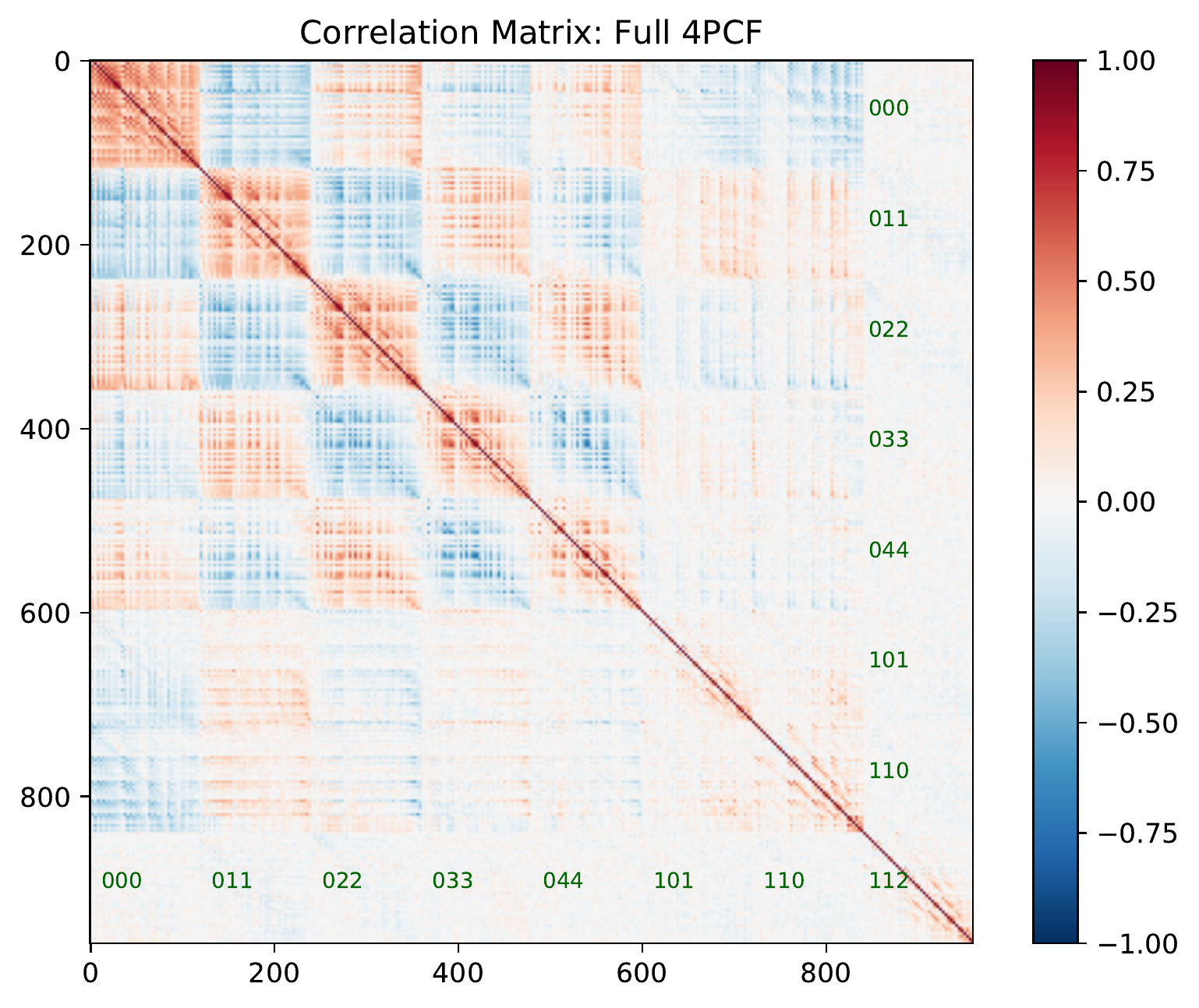}
    \includegraphics[width=0.48\textwidth]{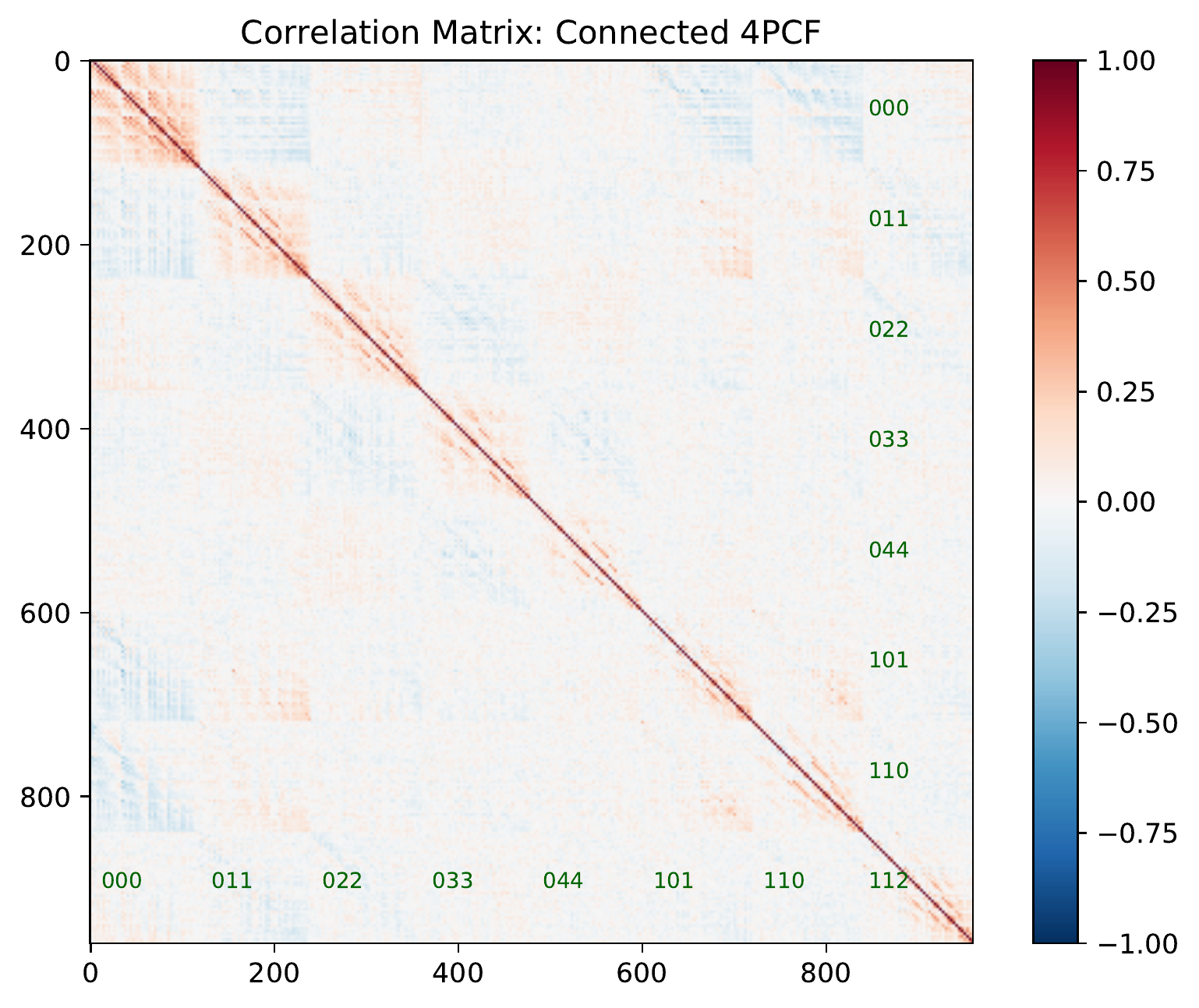}\\
    \caption{Correlation matrices for the full and connected 4PCFs, as measured from a suite of 1\,000 lognormal mocks at $z  = 2$. The connected 4PCF contribution is computed as the difference between full and disconnected 4PCFs, as in \eqref{eq: connected-4pcf-estimator}. Each $120\times120$ submatrix contains a single pair of multiplets with $\{\ell_1,\ell_2,\ell_3\}$ ($\{\ell'_1,\ell'_2,\ell'_3\}$) indicated in green on the bottom (right) of each matrix. We note that the 4PCF becomes substantially less correlated after subtracting the disconnected piece (which sources the partially-coupled covariance contributions discussed in \S\ref{subsec: covariance}).}
    \label{fig: lognormal-4pcf-corr}
\end{figure}

Fig.\,\ref{fig: lognormal-4pcf-corr} displays the empirical correlation matrices of the full and connected 4PCF (\textit{i.e.} $\mathrm{Cov}^{(\rm full, full)}$ and $\mathrm{Cov}^{( c, c)}$) obtained from the lognormal simulations of \S\ref{sec: lognormals}. Notably, removal of the disconnected terms leads to a \textit{far} more diagonal correlation matrix with significantly reduced bin-to-bin correlations, particularly between multiplets with one $\ell$ equal to zero (\textit{i.e.} those containing a Gaussian disconnected piece in the isotropic limit, cf.\,\ref{eq: iso-gaussian-4pcf}). The correlation is also significantly reduced compared to that of the Patchy mocks (Fig.\,\ref{fig: cov-comparison}) due to the higher redshift adopted. This illustrates the utility of subtracting the disconnected terms at the estimator level, rather than simply including them in the theory model using the results of Appendix \ref{appen: gaussian-deriv}.

\bibliographystyle{JHEP}
\bibliography{adslib,otherlib}

\end{document}